% File jytex.tex, for jyTeX version 2.6M (June 1992)
% Copyright (c) 1991, 1992 by Jonathan P. Yamron
% For full documentation, "get jydoc" from hep-ph@xxx.lanl.gov
%   Problems?  Contact brahm@theory3.caltech.edu.

\catcode`\@=11

%*****************************************************************************

\message{Loading jyTeX fonts...}

%************************************************************
%*
%*             Available fonts
%*
%************************************************************

%************** 5-point fonts *******************************

\font\vptrm=cmr5 \font\vptmit=cmmi5 \font\vptsy=cmsy5 \font\vptbf=cmbx5

\skewchar\vptmit='177 \skewchar\vptsy='60 \fontdimen16
\vptsy=\the\fontdimen17 \vptsy

\def\vpt{\ifmmode\err@badsizechange\else
     \@mathfontinit
     \textfont0=\vptrm  \scriptfont0=\vptrm  \scriptscriptfont0=\vptrm
     \textfont1=\vptmit \scriptfont1=\vptmit \scriptscriptfont1=\vptmit
     \textfont2=\vptsy  \scriptfont2=\vptsy  \scriptscriptfont2=\vptsy
     \textfont3=\xptex  \scriptfont3=\xptex  \scriptscriptfont3=\xptex
     \textfont\bffam=\vptbf
     \scriptfont\bffam=\vptbf
     \scriptscriptfont\bffam=\vptbf
     \@fontstyleinit
     \def\rm{\vptrm\fam=\z@}%
     \def\bf{\vptbf\fam=\bffam}%
     \def\oldstyle{\vptmit\fam=\@ne}%
     \rm\fi}

%************** 6-point fonts *******************************

\font\viptrm=cmr6 \font\viptmit=cmmi6 \font\viptsy=cmsy6
\font\viptbf=cmbx6

\skewchar\viptmit='177 \skewchar\viptsy='60 \fontdimen16
\viptsy=\the\fontdimen17 \viptsy

\def\vipt{\ifmmode\err@badsizechange\else
     \@mathfontinit
     \textfont0=\viptrm  \scriptfont0=\vptrm  \scriptscriptfont0=\vptrm
     \textfont1=\viptmit \scriptfont1=\vptmit \scriptscriptfont1=\vptmit
     \textfont2=\viptsy  \scriptfont2=\vptsy  \scriptscriptfont2=\vptsy
     \textfont3=\xptex   \scriptfont3=\xptex  \scriptscriptfont3=\xptex
     \textfont\bffam=\viptbf
     \scriptfont\bffam=\vptbf
     \scriptscriptfont\bffam=\vptbf
     \@fontstyleinit
     \def\rm{\viptrm\fam=\z@}%
     \def\bf{\viptbf\fam=\bffam}%
     \def\oldstyle{\viptmit\fam=\@ne}%
     \rm\fi}
%************** 7-point fonts *******************************

\font\viiptrm=cmr7 \font\viiptmit=cmmi7 \font\viiptsy=cmsy7
\font\viiptit=cmti7 \font\viiptbf=cmbx7

\skewchar\viiptmit='177 \skewchar\viiptsy='60 \fontdimen16
\viiptsy=\the\fontdimen17 \viiptsy

\def\viipt{\ifmmode\err@badsizechange\else
     \@mathfontinit
     \textfont0=\viiptrm  \scriptfont0=\vptrm  \scriptscriptfont0=\vptrm
     \textfont1=\viiptmit \scriptfont1=\vptmit \scriptscriptfont1=\vptmit
     \textfont2=\viiptsy  \scriptfont2=\vptsy  \scriptscriptfont2=\vptsy
     \textfont3=\xptex    \scriptfont3=\xptex  \scriptscriptfont3=\xptex
     \textfont\itfam=\viiptit
     \scriptfont\itfam=\viiptit
     \scriptscriptfont\itfam=\viiptit
     \textfont\bffam=\viiptbf
     \scriptfont\bffam=\vptbf
     \scriptscriptfont\bffam=\vptbf
     \@fontstyleinit
     \def\rm{\viiptrm\fam=\z@}%
     \def\it{\viiptit\fam=\itfam}%
     \def\bf{\viiptbf\fam=\bffam}%
     \def\oldstyle{\viiptmit\fam=\@ne}%
     \rm\fi}

%************** 8-point fonts *******************************

\font\viiiptrm=cmr8 \font\viiiptmit=cmmi8 \font\viiiptsy=cmsy8
\font\viiiptit=cmti8
%\font\viiiptsl=cmsl8
\font\viiiptbf=cmbx8
%\font\viiipttt=cmtt8
%\font\viiiptss=cmss8

\skewchar\viiiptmit='177 \skewchar\viiiptsy='60 \fontdimen16
\viiiptsy=\the\fontdimen17 \viiiptsy

\def\viiipt{\ifmmode\err@badsizechange\else
     \@mathfontinit
     \textfont0=\viiiptrm  \scriptfont0=\viptrm  \scriptscriptfont0=\vptrm
     \textfont1=\viiiptmit \scriptfont1=\viptmit \scriptscriptfont1=\vptmit
     \textfont2=\viiiptsy  \scriptfont2=\viptsy  \scriptscriptfont2=\vptsy
     \textfont3=\xptex     \scriptfont3=\xptex   \scriptscriptfont3=\xptex
     \textfont\itfam=\viiiptit
     \scriptfont\itfam=\viiptit
     \scriptscriptfont\itfam=\viiptit
     \textfont\bffam=\viiiptbf
     \scriptfont\bffam=\viptbf
     \scriptscriptfont\bffam=\vptbf
     \@fontstyleinit
     \def\rm{\viiiptrm\fam=\z@}%
     \def\it{\viiiptit\fam=\itfam}%
     \def\bf{\viiiptbf\fam=\bffam}%
     \def\oldstyle{\viiiptmit\fam=\@ne}%
     \rm\fi}

%************** Optional 9-point fonts **********************

\def\getixpt{%
     \font\ixptrm=cmr9
     \font\ixptmit=cmmi9
     \font\ixptsy=cmsy9
     \font\ixptit=cmti9
%     \font\ixptsl=cmsl9
     \font\ixptbf=cmbx9
%     \font\ixpttt=cmtt9
%     \font\ixptss=cmss9
     \skewchar\ixptmit='177 \skewchar\ixptsy='60
     \fontdimen16 \ixptsy=\the\fontdimen17 \ixptsy}

\def\ixpt{\ifmmode\err@badsizechange\else
     \@mathfontinit
     \textfont0=\ixptrm  \scriptfont0=\viiptrm  \scriptscriptfont0=\vptrm
     \textfont1=\ixptmit \scriptfont1=\viiptmit \scriptscriptfont1=\vptmit
     \textfont2=\ixptsy  \scriptfont2=\viiptsy  \scriptscriptfont2=\vptsy
     \textfont3=\xptex   \scriptfont3=\xptex    \scriptscriptfont3=\xptex
     \textfont\itfam=\ixptit
     \scriptfont\itfam=\viiptit
     \scriptscriptfont\itfam=\viiptit
     \textfont\bffam=\ixptbf
     \scriptfont\bffam=\viiptbf
     \scriptscriptfont\bffam=\vptbf
     \@fontstyleinit
     \def\rm{\ixptrm\fam=\z@}%
     \def\it{\ixptit\fam=\itfam}%
     \def\bf{\ixptbf\fam=\bffam}%
     \def\oldstyle{\ixptmit\fam=\@ne}%
     \rm\fi}

%************** 10-point fonts ******************************

\font\xptrm=cmr10 \font\xptmit=cmmi10 \font\xptsy=cmsy10
\font\xptex=cmex10 \font\xptit=cmti10 \font\xptsl=cmsl10
\font\xptbf=cmbx10 \font\xpttt=cmtt10 \font\xptss=cmss10
\font\xptsc=cmcsc10 \font\xptbfs=cmb10 \font\xptbmit=cmmib10

\skewchar\xptmit='177 \skewchar\xptbmit='177 \skewchar\xptsy='60
\fontdimen16 \xptsy=\the\fontdimen17 \xptsy

\def\xpt{\ifmmode\err@badsizechange\else
     \@mathfontinit
     \textfont0=\xptrm  \scriptfont0=\viiptrm  \scriptscriptfont0=\vptrm
     \textfont1=\xptmit \scriptfont1=\viiptmit \scriptscriptfont1=\vptmit
     \textfont2=\xptsy  \scriptfont2=\viiptsy  \scriptscriptfont2=\vptsy
     \textfont3=\xptex  \scriptfont3=\xptex    \scriptscriptfont3=\xptex
     \textfont\itfam=\xptit
     \scriptfont\itfam=\viiptit
     \scriptscriptfont\itfam=\viiptit
     \textfont\bffam=\xptbf
     \scriptfont\bffam=\viiptbf
     \scriptscriptfont\bffam=\vptbf
     \textfont\bfsfam=\xptbfs
     \scriptfont\bfsfam=\viiptbf
     \scriptscriptfont\bfsfam=\vptbf
     \textfont\bmitfam=\xptbmit
     \scriptfont\bmitfam=\viiptmit
     \scriptscriptfont\bmitfam=\vptmit
     \@fontstyleinit
     \def\rm{\xptrm\fam=\z@}%
     \def\it{\xptit\fam=\itfam}%
     \def\sl{\xptsl}%
     \def\bf{\xptbf\fam=\bffam}%
     \def\tt{\xpttt}%
     \def\ss{\xptss}%
     \def\sc{\xptsc}%
     \def\bfs{\xptbfs\fam=\bfsfam}%
     \def\bmit{\fam=\bmitfam}%
     \def\oldstyle{\xptmit\fam=\@ne}%
     \rm\fi}

%************** Optional 11-point fonts *********************

\def\getxipt{%
     \font\xiptrm=cmr10  scaled\magstephalf
     \font\xiptmit=cmmi10 scaled\magstephalf
     \font\xiptsy=cmsy10 scaled\magstephalf
     \font\xiptex=cmex10 scaled\magstephalf
     \font\xiptit=cmti10 scaled\magstephalf
     \font\xiptsl=cmsl10 scaled\magstephalf
     \font\xiptbf=cmbx10 scaled\magstephalf
     \font\xipttt=cmtt10 scaled\magstephalf
     \font\xiptss=cmss10 scaled\magstephalf
     \skewchar\xiptmit='177 \skewchar\xiptsy='60
     \fontdimen16 \xiptsy=\the\fontdimen17 \xiptsy}

\def\xipt{\ifmmode\err@badsizechange\else
     \@mathfontinit
     \textfont0=\xiptrm  \scriptfont0=\viiiptrm  \scriptscriptfont0=\viptrm
     \textfont1=\xiptmit \scriptfont1=\viiiptmit \scriptscriptfont1=\viptmit
     \textfont2=\xiptsy  \scriptfont2=\viiiptsy  \scriptscriptfont2=\viptsy
     \textfont3=\xiptex  \scriptfont3=\xptex     \scriptscriptfont3=\xptex
     \textfont\itfam=\xiptit
     \scriptfont\itfam=\viiiptit
     \scriptscriptfont\itfam=\viiptit
     \textfont\bffam=\xiptbf
     \scriptfont\bffam=\viiiptbf
     \scriptscriptfont\bffam=\viptbf
     \@fontstyleinit
     \def\rm{\xiptrm\fam=\z@}%
     \def\it{\xiptit\fam=\itfam}%
     \def\sl{\xiptsl}%
     \def\bf{\xiptbf\fam=\bffam}%
     \def\tt{\xipttt}%
     \def\ss{\xiptss}%
     \def\oldstyle{\xiptmit\fam=\@ne}%
     \rm\fi}

%************** 12-point fonts ******************************

\font\xiiptrm=cmr12 \font\xiiptmit=cmmi12 \font\xiiptsy=cmsy10
scaled\magstep1 \font\xiiptex=cmex10  scaled\magstep1
\font\xiiptit=cmti12 \font\xiiptsl=cmsl12 \font\xiiptbf=cmbx12
%\font\xiipttt=cmtt12
\font\xiiptss=cmss12 \font\xiiptsc=cmcsc10 scaled\magstep1
\font\xiiptbfs=cmb10  scaled\magstep1 \font\xiiptbmit=cmmib10
scaled\magstep1

\skewchar\xiiptmit='177 \skewchar\xiiptbmit='177 \skewchar\xiiptsy='60
\fontdimen16 \xiiptsy=\the\fontdimen17 \xiiptsy

\def\xiipt{\ifmmode\err@badsizechange\else
     \@mathfontinit
     \textfont0=\xiiptrm  \scriptfont0=\viiiptrm  \scriptscriptfont0=\viptrm
     \textfont1=\xiiptmit \scriptfont1=\viiiptmit \scriptscriptfont1=\viptmit
     \textfont2=\xiiptsy  \scriptfont2=\viiiptsy  \scriptscriptfont2=\viptsy
     \textfont3=\xiiptex  \scriptfont3=\xptex     \scriptscriptfont3=\xptex
     \textfont\itfam=\xiiptit
     \scriptfont\itfam=\viiiptit
     \scriptscriptfont\itfam=\viiptit
     \textfont\bffam=\xiiptbf
     \scriptfont\bffam=\viiiptbf
     \scriptscriptfont\bffam=\viptbf
     \textfont\bfsfam=\xiiptbfs
     \scriptfont\bfsfam=\viiiptbf
     \scriptscriptfont\bfsfam=\viptbf
     \textfont\bmitfam=\xiiptbmit
     \scriptfont\bmitfam=\viiiptmit
     \scriptscriptfont\bmitfam=\viptmit
     \@fontstyleinit
     \def\rm{\xiiptrm\fam=\z@}%
     \def\it{\xiiptit\fam=\itfam}%
     \def\sl{\xiiptsl}%
     \def\bf{\xiiptbf\fam=\bffam}%
     \def\tt{\xiipttt}%
     \def\ss{\xiiptss}%
     \def\sc{\xiiptsc}%
     \def\bfs{\xiiptbfs\fam=\bfsfam}%
     \def\bmit{\fam=\bmitfam}%
     \def\oldstyle{\xiiptmit\fam=\@ne}%
     \rm\fi}

%************** Optional 13-point fonts *********************

\def\getxiiipt{%
     \font\xiiiptrm=cmr12  scaled\magstephalf
     \font\xiiiptmit=cmmi12 scaled\magstephalf
     \font\xiiiptsy=cmsy9  scaled\magstep2
     \font\xiiiptit=cmti12 scaled\magstephalf
     \font\xiiiptsl=cmsl12 scaled\magstephalf
     \font\xiiiptbf=cmbx12 scaled\magstephalf
     \font\xiiipttt=cmtt12 scaled\magstephalf
     \font\xiiiptss=cmss12 scaled\magstephalf
     \skewchar\xiiiptmit='177 \skewchar\xiiiptsy='60
     \fontdimen16 \xiiiptsy=\the\fontdimen17 \xiiiptsy}

\def\xiiipt{\ifmmode\err@badsizechange\else
     \@mathfontinit
     \textfont0=\xiiiptrm  \scriptfont0=\xptrm  \scriptscriptfont0=\viiptrm
     \textfont1=\xiiiptmit \scriptfont1=\xptmit \scriptscriptfont1=\viiptmit
     \textfont2=\xiiiptsy  \scriptfont2=\xptsy  \scriptscriptfont2=\viiptsy
     \textfont3=\xivptex   \scriptfont3=\xptex  \scriptscriptfont3=\xptex
     \textfont\itfam=\xiiiptit
     \scriptfont\itfam=\xptit
     \scriptscriptfont\itfam=\viiptit
     \textfont\bffam=\xiiiptbf
     \scriptfont\bffam=\xptbf
     \scriptscriptfont\bffam=\viiptbf
     \@fontstyleinit
     \def\rm{\xiiiptrm\fam=\z@}%
     \def\it{\xiiiptit\fam=\itfam}%
     \def\sl{\xiiiptsl}%
     \def\bf{\xiiiptbf\fam=\bffam}%
     \def\tt{\xiiipttt}%
     \def\ss{\xiiiptss}%
     \def\oldstyle{\xiiiptmit\fam=\@ne}%
     \rm\fi}

%************** 14-point fonts ******************************

\font\xivptrm=cmr12   scaled\magstep1 \font\xivptmit=cmmi12
scaled\magstep1 \font\xivptsy=cmsy10  scaled\magstep2
\font\xivptex=cmex10  scaled\magstep2 \font\xivptit=cmti12
scaled\magstep1 \font\xivptsl=cmsl12  scaled\magstep1
\font\xivptbf=cmbx12  scaled\magstep1
%\font\xivpttt=cmtt12  scaled\magstep1
\font\xivptss=cmss12  scaled\magstep1 \font\xivptsc=cmcsc10
scaled\magstep2 \font\xivptbfs=cmb10  scaled\magstep2
\font\xivptbmit=cmmib10 scaled\magstep2

\skewchar\xivptmit='177 \skewchar\xivptbmit='177 \skewchar\xivptsy='60
\fontdimen16 \xivptsy=\the\fontdimen17 \xivptsy

\def\xivpt{\ifmmode\err@badsizechange\else
     \@mathfontinit
     \textfont0=\xivptrm  \scriptfont0=\xptrm  \scriptscriptfont0=\viiptrm
     \textfont1=\xivptmit \scriptfont1=\xptmit \scriptscriptfont1=\viiptmit
     \textfont2=\xivptsy  \scriptfont2=\xptsy  \scriptscriptfont2=\viiptsy
     \textfont3=\xivptex  \scriptfont3=\xptex  \scriptscriptfont3=\xptex
     \textfont\itfam=\xivptit
     \scriptfont\itfam=\xptit
     \scriptscriptfont\itfam=\viiptit
     \textfont\bffam=\xivptbf
     \scriptfont\bffam=\xptbf
     \scriptscriptfont\bffam=\viiptbf
     \textfont\bfsfam=\xivptbfs
     \scriptfont\bfsfam=\xptbfs
     \scriptscriptfont\bfsfam=\viiptbf
     \textfont\bmitfam=\xivptbmit
     \scriptfont\bmitfam=\xptbmit
     \scriptscriptfont\bmitfam=\viiptmit
     \@fontstyleinit
     \def\rm{\xivptrm\fam=\z@}%
     \def\it{\xivptit\fam=\itfam}%
     \def\sl{\xivptsl}%
     \def\bf{\xivptbf\fam=\bffam}%
     \def\tt{\xivpttt}%
     \def\ss{\xivptss}%
     \def\sc{\xivptsc}%
     \def\bfs{\xivptbfs\fam=\bfsfam}%
     \def\bmit{\fam=\bmitfam}%
     \def\oldstyle{\xivptmit\fam=\@ne}%
     \rm\fi}

%************** 17-point fonts ******************************

\font\xviiptrm=cmr17 \font\xviiptmit=cmmi12 scaled\magstep2
\font\xviiptsy=cmsy10 scaled\magstep3 \font\xviiptex=cmex10
scaled\magstep3 \font\xviiptit=cmti12 scaled\magstep2
\font\xviiptbf=cmbx12 scaled\magstep2 \font\xviiptbfs=cmb10
scaled\magstep3

\skewchar\xviiptmit='177 \skewchar\xviiptsy='60 \fontdimen16
\xviiptsy=\the\fontdimen17 \xviiptsy

\def\xviipt{\ifmmode\err@badsizechange\else
     \@mathfontinit
     \textfont0=\xviiptrm  \scriptfont0=\xiiptrm  \scriptscriptfont0=\viiiptrm
     \textfont1=\xviiptmit \scriptfont1=\xiiptmit \scriptscriptfont1=\viiiptmit
     \textfont2=\xviiptsy  \scriptfont2=\xiiptsy  \scriptscriptfont2=\viiiptsy
     \textfont3=\xviiptex  \scriptfont3=\xiiptex  \scriptscriptfont3=\xptex
     \textfont\itfam=\xviiptit
     \scriptfont\itfam=\xiiptit
     \scriptscriptfont\itfam=\viiiptit
     \textfont\bffam=\xviiptbf
     \scriptfont\bffam=\xiiptbf
     \scriptscriptfont\bffam=\viiiptbf
     \textfont\bfsfam=\xviiptbfs
     \scriptfont\bfsfam=\xiiptbfs
     \scriptscriptfont\bfsfam=\viiiptbf
     \@fontstyleinit
     \def\rm{\xviiptrm\fam=\z@}%
     \def\it{\xviiptit\fam=\itfam}%
     \def\bf{\xviiptbf\fam=\bffam}%
     \def\bfs{\xviiptbfs\fam=\bfsfam}%
     \def\oldstyle{\xviiptmit\fam=\@ne}%
     \rm\fi}

%************** 21-point fonts ******************************

\font\xxiptrm=cmr17  scaled\magstep1
%\font\xxiptmit=cmmi12 scaled\magstep3
%\font\xxiptsy=cmsy10 scaled\magstep4
%\font\xxiptex=cmex10 scaled\magstep4
%\font\xxiptbf=cmbx12 scaled\magstep3

%\skewchar\xxiptmit='177 \skewchar\xxiptsy='60
%\fontdimen16 \xxiptsy=\the\fontdimen17 \xxiptsy

\def\xxipt{\ifmmode\err@badsizechange\else
     \@mathfontinit
%     \textfont0=\xxiptrm  \scriptfont0=\xivptrm  \scriptscriptfont0=\xptrm
%     \textfont1=\xxiptmit \scriptfont1=\xivptmit \scriptscriptfont1=\xptmit
%     \textfont2=\xxiptsy  \scriptfont2=\xivptsy  \scriptscriptfont2=\xptsy
%     \textfont3=\xxiptex  \scriptfont3=\xivptex  \scriptscriptfont3=\xptex
%     \textfont\bffam=\xxiptbf
%     \scriptfont\bffam=\xivptbf
%     \scriptscriptfont\bffam=\xptbf
     \@fontstyleinit
     \def\rm{\xxiptrm\fam=\z@}%
     \rm\fi}

%************** 25-point fonts ******************************

\font\xxvptrm=cmr17  scaled\magstep2
%\font\xxvptmit=cmmi12 scaled\magstep4
%\font\xxvptsy=cmsy10 scaled\magstep5
%\font\xxvptex=cmex10 scaled\magstep5
%\font\xxvptbf=cmbx12 scaled\magstep4

%\skewchar\xxvptmit='177 \skewchar\xxvptsy='60
%\fontdimen16 \xxvptsy=\the\fontdimen17 \xxvptsy

\def\xxvpt{\ifmmode\err@badsizechange\else
     \@mathfontinit
%     \textfont0=\xxvptrm  \scriptfont0=\xviiptrm  \scriptscriptfont0=\xiiptrm
%     \textfont1=\xxvptmit \scriptfont1=\xviiptmit \scriptscriptfont1=\xiiptmit
%     \textfont2=\xxvptsy  \scriptfont2=\xviiptsy  \scriptscriptfont2=\xiiptsy
%     \textfont3=\xxvptex  \scriptfont3=\xviiptex  \scriptscriptfont3=\xiiptex
%     \textfont\bffam=\xxvptbf
%     \scriptfont\bffam=\xviiptbf
%     \scriptscriptfont\bffam=\xiiptbf
     \@fontstyleinit
     \def\rm{\xxvptrm\fam=\z@}%
     \rm\fi}

%************** Other fonts *********************************

%\font\dummy=dummy

%******************************************************************************

\message{Loading jyTeX macros...}

%************************************************************
%*
%*              Simple modifications to plain
%*
%************************************************************
\message{modifications to plain.tex,}

% The "\outer" qualifier is removed from the definitions of \newcount through
% \newif so that they may be used in definitions.  \newif is also changed to
% make \if commands globally defined.

\def\newcount{\alloc@0\count\countdef\insc@unt}
\def\newdimen{\alloc@1\dimen\dimendef\insc@unt}
\def\newskip{\alloc@2\skip\skipdef\insc@unt}
\def\newmuskip{\alloc@3\muskip\muskipdef\@cclvi}
\def\newbox{\alloc@4\box\chardef\insc@unt}
\def\newtoks{\alloc@5\toks\toksdef\@cclvi}
\def\newhelp#1#2{\newtoks#1\global#1\expandafter{\csname#2\endcsname}}
\def\newread{\alloc@6\read\chardef\sixt@@n}
\def\newwrite{\alloc@7\write\chardef\sixt@@n}
\def\newfam{\alloc@8\fam\chardef\sixt@@n}
\def\newinsert#1{\global\advance\insc@unt by\m@ne
     \ch@ck0\insc@unt\count
     \ch@ck1\insc@unt\dimen
     \ch@ck2\insc@unt\skip
     \ch@ck4\insc@unt\box
     \allocationnumber=\insc@unt
     \global\chardef#1=\allocationnumber
     \wlog{\string#1=\string\insert\the\allocationnumber}}
\def\newif#1{\count@\escapechar \escapechar\m@ne
     \expandafter\expandafter\expandafter
          \xdef\@if#1{true}{\let\noexpand#1=\noexpand\iftrue}%
     \expandafter\expandafter\expandafter
          \xdef\@if#1{false}{\let\noexpand#1=\noexpand\iffalse}%
     \global\@if#1{false}\escapechar=\count@}

%************** Some parameter changes **********************

\newlinechar=`\^^J
\overfullrule=0pt

%************** Font-related modifications ******************

% The plain fonts are mapped onto the corresponding jyTeX fonts

% Some control sequences are disabled.

\let\itfam=\undefined

\let\bffam=\undefined

\count18=3

% German sharp s is given a new name (\ss is already taken)

\chardef\sharps="19

% The mathcode assignments of characters in the math italic font are changed to
% allow for switching to boldface.

\mathchardef\alpha="710B \mathchardef\beta="710C \mathchardef\gamma="710D
\mathchardef\delta="710E \mathchardef\epsilon="710F
\mathchardef\zeta="7110 \mathchardef\eta="7111 \mathchardef\theta="7112
\mathchardef\iota="7113 \mathchardef\kappa="7114
\mathchardef\lambda="7115 \mathchardef\mu="7116 \mathchardef\nu="7117
\mathchardef\xi="7118 \mathchardef\pi="7119 \mathchardef\rho="711A
\mathchardef\sigma="711B \mathchardef\tau="711C
\mathchardef\upsilon="711D \mathchardef\phi="711E \mathchardef\chi="711F
\mathchardef\psi="7120 \mathchardef\omega="7121
\mathchardef\varepsilon="7122 \mathchardef\vartheta="7123
\mathchardef\varpi="7124 \mathchardef\varrho="7125
\mathchardef\varsigma="7126 \mathchardef\varphi="7127
\mathchardef\imath="717B \mathchardef\jmath="717C \mathchardef\ell="7160
\mathchardef\wp="717D \mathchardef\partial="7140 \mathchardef\flat="715B
\mathchardef\natural="715C \mathchardef\sharp="715D

%************** Miscellaneous changes ***********************

% The dimension \p@ (1pt) is replaced with \rp@ (relative pt, defined below),
% whose size is determined by the base type size of the document.

\def\angle{{\vbox{\ialign{$\m@th\scriptstyle##$\crcr
     \not\mathrel{\mkern14mu}\crcr
     \noalign{\nointerlineskip}
     \mkern2.5mu\leaders\hrule height.34\rp@\hfill\mkern2.5mu\crcr}}}}
\def\vdots{\vbox{\baselineskip4\rp@ \lineskiplimit\z@
     \kern6\rp@\hbox{.}\hbox{.}\hbox{.}}}
\def\ddots{\mathinner{\mkern1mu\raise7\rp@\vbox{\kern7\rp@\hbox{.}}\mkern2mu
     \raise4\rp@\hbox{.}\mkern2mu\raise\rp@\hbox{.}\mkern1mu}}
\def\overrightarrow#1{\vbox{\ialign{##\crcr
     \rightarrowfill\crcr
     \noalign{\kern-\rp@\nointerlineskip}
     $\hfil\displaystyle{#1}\hfil$\crcr}}}
\def\overleftarrow#1{\vbox{\ialign{##\crcr
     \leftarrowfill\crcr
     \noalign{\kern-\rp@\nointerlineskip}
     $\hfil\displaystyle{#1}\hfil$\crcr}}}
\def\overbrace#1{\mathop{\vbox{\ialign{##\crcr
     \noalign{\kern3\rp@}
     \downbracefill\crcr
     \noalign{\kern3\rp@\nointerlineskip}
     $\hfil\displaystyle{#1}\hfil$\crcr}}}\limits}
\def\underbrace#1{\mathop{\vtop{\ialign{##\crcr
     $\hfil\displaystyle{#1}\hfil$\crcr
     \noalign{\kern3\rp@\nointerlineskip}
     \upbracefill\crcr
     \noalign{\kern3\rp@}}}}\limits}
\def\big#1{{\hbox{$\left#1\vbox to8.5\rp@ {}\right.\n@space$}}}
\def\Big#1{{\hbox{$\left#1\vbox to11.5\rp@ {}\right.\n@space$}}}
\def\bigg#1{{\hbox{$\left#1\vbox to14.5\rp@ {}\right.\n@space$}}}
\def\Bigg#1{{\hbox{$\left#1\vbox to17.5\rp@ {}\right.\n@space$}}}
\def\@vereq#1#2{\lower.5\rp@\vbox{\baselineskip\z@skip\lineskip-.5\rp@
     \ialign{$\m@th#1\hfil##\hfil$\crcr#2\crcr=\crcr}}}
\def\rlh@#1{\vcenter{\hbox{\ooalign{\raise2\rp@
     \hbox{$#1\rightharpoonup$}\crcr
     $#1\leftharpoondown$}}}}
\def\bordermatrix#1{\begingroup\m@th
     \setbox\z@\vbox{%
          \def\cr{\crcr\noalign{\kern2\rp@\global\let\cr\endline}}%
          \ialign{$##$\hfil\kern2\rp@\kern\p@renwd
               &\thinspace\hfil$##$\hfil&&\quad\hfil$##$\hfil\crcr
               \omit\strut\hfil\crcr
               \noalign{\kern-\baselineskip}%
               #1\crcr\omit\strut\cr}}%
     \setbox\tw@\vbox{\unvcopy\z@\global\setbox\@ne\lastbox}%
     \setbox\tw@\hbox{\unhbox\@ne\unskip\global\setbox\@ne\lastbox}%
     \setbox\tw@\hbox{$\kern\wd\@ne\kern-\p@renwd\left(\kern-\wd\@ne
          \global\setbox\@ne\vbox{\box\@ne\kern2\rp@}%
          \vcenter{\kern-\ht\@ne\unvbox\z@\kern-\baselineskip}%
          \,\right)$}%
     \null\;\vbox{\kern\ht\@ne\box\tw@}\endgroup}
\def\endinsert{\egroup
     \if@mid\dimen@\ht\z@
          \advance\dimen@\dp\z@
          \advance\dimen@12\rp@
          \advance\dimen@\pagetotal
          \ifdim\dimen@>\pagegoal\@midfalse\p@gefalse\fi
     \fi
     \if@mid\bigskip\box\z@
          \bigbreak
     \else\insert\topins{\penalty100 \splittopskip\z@skip
               \splitmaxdepth\maxdimen\floatingpenalty\z@
               \ifp@ge\dimen@\dp\z@
                    \vbox to\vsize{\unvbox\z@\kern-\dimen@}%
               \else\box\z@\nobreak\bigskip
               \fi}%
     \fi
     \endgroup}

% \normalbaselines is removed from \cases and \matrix.

\def\cases#1{\left\{\,\vcenter{\m@th
     \ialign{$##\hfil$&\quad##\hfil\crcr#1\crcr}}\right.}
\def\matrix#1{\null\,\vcenter{\m@th
     \ialign{\hfil$##$\hfil&&\quad\hfil$##$\hfil\crcr
          \mathstrut\crcr
          \noalign{\kern-\baselineskip}
          #1\crcr
          \mathstrut\crcr
          \noalign{\kern-\baselineskip}}}\,}

% \raggedbottom modified slightly

\newif\ifraggedbottom

\def\raggedbottom{\ifraggedbottom\else
     \advance\topskip by\z@ plus60pt \raggedbottomtrue\fi}%
\def\normalbottom{\ifraggedbottom
     \advance\topskip by\z@ plus-60pt \raggedbottomfalse\fi}

%************************************************************
%*
%*              Miscellaneous definitions
%*
%************************************************************
\message{hacks,}

%************** Hack registers ******************************

\toksdef\toks@i=1 \toksdef\toks@ii=2

%************** Basic macros ********************************

\def\TeX{T\kern-.1667em \lower.5ex \hbox{E}\kern-.125em X\null}
\def\jyTeX{{\leavevmode
     \raise.587ex \hbox{\it\j}\kern-.1em \lower.048ex \hbox{\it y}\kern-.12em
     \TeX}}

\let\then=\iftrue
\def\ifnoarg#1\then{\def\hack@{#1}\ifx\hack@\empty}
\def\ifundefined#1\then{%
     \expandafter\ifx\csname\expandafter\blank\string#1\endcsname\relax}
\def\useif#1\then{\csname#1\endcsname}
\def\usename#1{\csname#1\endcsname}
\def\useafter#1#2{\expandafter#1\csname#2\endcsname}

% Modify so that I can have \loop's within \loop's?
\long\def\loop#1\repeat{\def\@iterate{#1\expandafter\@iterate\fi}\@iterate
     \let\@iterate=\relax}
%\long\def\loop#1\repeat{\def\@loopbody{#1}\@iterate}
%\def\@iterate{\@loopbody\let\next=\@iterate\else\let\next=\relax\fi\next}

\let\TeXend=\end
\def\begin#1{\begingroup\def\@@blockname{#1}\usename{begin#1}}
\def\end#1{\usename{end#1}\def\hack@{#1}%
     \ifx\@@blockname\hack@
          \endgroup
     \else\err@badgroup\hack@\@@blockname
     \fi}
\def\@@blockname{}

\def\defaultoption[#1]#2{%
     \def\hack@{\ifx\hack@ii[\toks@={#2}\else\toks@={#2[#1]}\fi\the\toks@}%
     \futurelet\hack@ii\hack@}

\def\markup#1{\let\@@marksf=\empty
     \ifhmode\edef\@@marksf{\spacefactor=\the\spacefactor\relax}\/\fi
     ${}^{\hbox{\subscriptfonts#1}}$\@@marksf}

%************** Time registers ******************************

\newtoks\shortyear
\newtoks\militaryhour
\newtoks\standardhour
\newtoks\minute
\newtoks\amorpm

\def\settime{\count@=\time\divide\count@ by60
     \militaryhour=\expandafter{\number\count@}%
     {\multiply\count@ by-60 \advance\count@ by\time
          \xdef\hack@{\ifnum\count@<10 0\fi\number\count@}}%
     \minute=\expandafter{\hack@}%
     \ifnum\count@<12
          \amorpm={am}
     \else\amorpm={pm}
          \ifnum\count@>12 \advance\count@ by-12 \fi
     \fi
     \standardhour=\expandafter{\number\count@}%
     \def\hack@19##1##2{\shortyear={##1##2}}%
          \expandafter\hack@\the\year}

\def\monthword#1{%
     \ifcase#1
          $\bullet$\err@badcountervalue{monthword}%
          \or January\or February\or March\or April\or May\or June%
          \or July\or August\or September\or October\or November\or December%
     \else$\bullet$\err@badcountervalue{monthword}%
     \fi}

\def\monthabbr#1{%
     \ifcase#1
          $\bullet$\err@badcountervalue{monthabbr}%
          \or Jan\or Feb\or Mar\or Apr\or May\or Jun%
          \or Jul\or Aug\or Sep\or Oct\or Nov\or Dec%
     \else$\bullet$\err@badcountervalue{monthabbr}%
     \fi}

\def\militarytime{\the\militaryhour:\the\minute}
\def\standardtime{\the\standardhour:\the\minute}

%************** Number styles *******************************

\def\@setnumstyle#1#2{\expandafter\global\expandafter\expandafter
     \expandafter\let\expandafter\expandafter
     \csname @\expandafter\blank\string#1style\endcsname
     \csname#2\endcsname}
\def\numstyle#1{\usename{@\expandafter\blank\string#1style}#1}
\def\ifblank#1\then{\useafter\ifx{@\expandafter\blank\string#1}\blank}

\def\blank#1{}

\def\Roman#1{\expandafter\uppercase\expandafter{\romannumeral#1}}
\def\alphabetic#1{%
     \ifcase#1
          $\bullet$\err@badcountervalue{alphabetic}%
          \or a\or b\or c\or d\or e\or f\or g\or h\or i\or j\or k\or l\or m%
          \or n\or o\or p\or q\or r\or s\or t\or u\or v\or w\or x\or y\or z%
     \else$\bullet$\err@badcountervalue{alphabetic}%
     \fi}
\def\Alphabetic#1{\expandafter\uppercase\expandafter{\alphabetic{#1}}}
\def\symbols#1{%
     \ifcase#1
          $\bullet$\err@badcountervalue{symbols}%
          \or*\or\dag\or\ddag\or\S\or$\|$%
          \or**\or\dag\dag\or\ddag\ddag\or\S\S\or$\|\|$%
     \else$\bullet$\err@badcountervalue{symbols}%
     \fi}

%************** String macros *******************************

\catcode`\^^?=13 \def^^?{\relax}

\def\trimleading#1\to#2{\edef#2{#1}%
     \expandafter\@trimleading\expandafter#2#2^^?^^?}
\def\@trimleading#1#2#3^^?{\ifx#2^^?\def#1{}\else\def#1{#2#3}\fi}

\def\trimtrailing#1\to#2{\edef#2{#1}%
     \expandafter\@trimtrailing\expandafter#2#2^^? ^^?\relax}
\def\@trimtrailing#1#2 ^^?#3{\ifx#3\relax\toks@={}%
     \else\def#1{#2}\toks@={\trimtrailing#1\to#1}\fi
     \the\toks@}

\def\trim#1\to#2{\trimleading#1\to#2\trimtrailing#2\to#2}

\catcode`\^^?=15

%************** List macros *********************************

\long\def\additemL#1\to#2{\toks@={\^^\{#1}}\toks@ii=\expandafter{#2}%
     \xdef#2{\the\toks@\the\toks@ii}}

\long\def\additemR#1\to#2{\toks@={\^^\{#1}}\toks@ii=\expandafter{#2}%
     \xdef#2{\the\toks@ii\the\toks@}}

\def\getitemL#1\to#2{\expandafter\@getitemL#1\hack@#1#2}
\def\@getitemL\^^\#1#2\hack@#3#4{\def#4{#1}\def#3{#2}}

%************************************************************
%*
%*             Font-related macros
%*
%************************************************************
\message{font macros,}

%************** Font set-up *********************************

\newdimen\rp@
\newcount\@@sizeindex \@@sizeindex=0
\newcount\@@factori
\newcount\@@factorii
\newcount\@@factoriii
\newcount\@@factoriv

\countdef\maxfam=18
\newfam\itfam
\newfam\bffam
\newfam\bfsfam
\newfam\bmitfam

\def\@mathfontinit{\count@=4
     \loop\textfont\count@=\nullfont
          \scriptfont\count@=\nullfont
          \scriptscriptfont\count@=\nullfont
          \ifnum\count@<\maxfam\advance\count@ by\@ne
     \repeat}

\def\@fontstyleinit{%
     \def\it{\err@fontnotavailable\it}%
     \def\bf{\err@fontnotavailable\bf}%
     \def\bfs{\err@bfstobf}%
     \def\bmit{\err@fontnotavailable\bmit}%
     \def\sc{\err@fontnotavailable\sc}%
     \def\sl{\err@sltoit}%
     \def\ss{\err@fontnotavailable\ss}%
     \def\tt{\err@fontnotavailable\tt}}

\def\@parameterinit#1{\rm\rp@=.1em \@getscaling{#1}%
     \let\^^\=\@doscaling\scalingskipslist
     \setbox\strutbox=\hbox{\vrule
          height.708\baselineskip depth.292\baselineskip width\z@}}

\def\@getfactor#1#2#3#4{\@@factori=#1 \@@factorii=#2
     \@@factoriii=#3 \@@factoriv=#4}

\def\@getscaling#1{\count@=#1 \advance\count@ by-\@@sizeindex\@@sizeindex=#1
     \ifnum\count@<0
          \let\@mulordiv=\divide
          \let\@divormul=\multiply
          \multiply\count@ by\m@ne
     \else\let\@mulordiv=\multiply
          \let\@divormul=\divide
     \fi
     \edef\@@scratcha{\ifcase\count@                {1}{1}{1}{1}\or
          {1}{7}{23}{3}\or     {2}{5}{3}{1}\or      {9}{89}{13}{1}\or
          {6}{25}{6}{1}\or     {8}{71}{14}{1}\or    {6}{25}{36}{5}\or
          {1}{7}{53}{4}\or     {12}{125}{108}{5}\or {3}{14}{53}{5}\or
          {6}{41}{17}{1}\or    {13}{31}{13}{2}\or   {9}{107}{71}{2}\or
          {11}{139}{124}{3}\or {1}{6}{43}{2}\or     {10}{107}{42}{1}\or
          {1}{5}{43}{2}\or     {5}{69}{65}{1}\or    {11}{97}{91}{2}\fi}%
     \expandafter\@getfactor\@@scratcha}

\def\@doscaling#1{\@mulordiv#1by\@@factori\@divormul#1by\@@factorii
     \@mulordiv#1by\@@factoriii\@divormul#1by\@@factoriv}

%************* Size-changing commands ***********************

\newskip\headskip
\newskip\footskip

\def\typesize=#1pt{\count@=#1 \advance\count@ by-10
     \ifcase\count@
          \@setsizex\or\err@badtypesize\or
          \@setsizexii\or\err@badtypesize\or
          \@setsizexiv
     \else\err@badtypesize
     \fi}

\def\@setsizex{\getixpt
     \def\subsubscriptfonts{\vpt}%
          \def\subsubscriptsize{\vpt\@parameterinit{-8}}%
     \def\subscriptfonts{\viipt}\def\subscriptsize{\viipt\@parameterinit{-4}}%
     \def\footnotefonts{\viiipt}\def\footnotesize{\viiipt\@parameterinit{-2}}%
     \def\smallfonts{\ixpt}\def\smallsize{\ixpt\@parameterinit{-1}}%
     \def\normalfonts{\xpt}\def\normalsize{\xpt\@parameterinit{0}}%
     \def\bigfonts{\xiipt}\def\bigsize{\xiipt\@parameterinit{2}}%
     \def\Bigfonts{\xivpt}\def\Bigsize{\xivpt\@parameterinit{4}}%
     \def\biggfonts{\xviipt}\def\biggsize{\xviipt\@parameterinit{6}}%
     \def\Biggfonts{\xxipt}\def\Biggsize{\xxipt\@parameterinit{8}}%
     \def\tinyfonts{\vpt}\def\tinysize{\vpt\@parameterinit{-8}}%
     \def\HUGEFONTS{\xxvpt}\def\HUGESIZE{\xxvpt\@parameterinit{10}}%
     \normalsize\fixedskipslist}

\def\@setsizexii{\getxipt
     \def\subsubscriptfonts{\vipt}%
          \def\subsubscriptsize{\vipt\@parameterinit{-6}}%
     \def\subscriptfonts{\viiipt}%
          \def\subscriptsize{\viiipt\@parameterinit{-2}}%
     \def\footnotefonts{\xpt}\def\footnotesize{\xpt\@parameterinit{0}}%
     \def\smallfonts{\xipt}\def\smallsize{\xipt\@parameterinit{1}}%
     \def\normalfonts{\xiipt}\def\normalsize{\xiipt\@parameterinit{2}}%
     \def\bigfonts{\xivpt}\def\bigsize{\xivpt\@parameterinit{4}}%
     \def\Bigfonts{\xviipt}\def\Bigsize{\xviipt\@parameterinit{6}}%
     \def\biggfonts{\xxipt}\def\biggsize{\xxipt\@parameterinit{8}}%
     \def\Biggfonts{\xxvpt}\def\Biggsize{\xxvpt\@parameterinit{10}}%
     \def\tinyfonts{\vpt}\def\tinysize{\vpt\@parameterinit{-8}}%
     \def\HUGEFONTS{\xxvpt}\def\HUGESIZE{\xxvpt\@parameterinit{10}}%
     \normalsize\fixedskipslist}

\def\@setsizexiv{\getxiiipt
     \def\subsubscriptfonts{\viipt}%
          \def\subsubscriptsize{\viipt\@parameterinit{-4}}%
     \def\subscriptfonts{\xpt}\def\subscriptsize{\xpt\@parameterinit{0}}%
     \def\footnotefonts{\xiipt}\def\footnotesize{\xiipt\@parameterinit{2}}%
     \def\smallfonts{\xiiipt}\def\smallsize{\xiiipt\@parameterinit{3}}%
     \def\normalfonts{\xivpt}\def\normalsize{\xivpt\@parameterinit{4}}%
     \def\bigfonts{\xviipt}\def\bigsize{\xviipt\@parameterinit{6}}%
     \def\Bigfonts{\xxipt}\def\Bigsize{\xxipt\@parameterinit{8}}%
     \def\biggfonts{\xxvpt}\def\biggsize{\xxvpt\@parameterinit{10}}%
     \def\Biggfonts{\err@sizetoolarge\Biggfonts\HUGEFONTS}%
          \def\Biggsize{\err@sizetoolarge\Biggsize\HUGESIZE}%
     \def\tinyfonts{\vpt}\def\tinysize{\vpt\@parameterinit{-8}}%
     \def\HUGEFONTS{\xxvpt}\def\HUGESIZE{\xxvpt\@parameterinit{10}}%
     \normalsize\fixedskipslist}

\def\subsubscriptfonts{\vpt} \def\subsubscriptsize{\vpt\@parameterinit{-8}}
\def\subscriptfonts{\viipt}  \def\subscriptsize{\viipt\@parameterinit{-4}}
\def\footnotefonts{\viiipt}  \def\footnotesize{\viiipt\@parameterinit{-2}}
\def\smallfonts{\err@sizenotavailable\smallfonts}
                             \def\smallsize{\ixpt\@parameterinit{-1}}
\def\normalfonts{\xpt}       \def\normalsize{\xpt\@parameterinit{0}}
\def\bigfonts{\xiipt}        \def\bigsize{\xiipt\@parameterinit{2}}
\def\Bigfonts{\xivpt}        \def\Bigsize{\xivpt\@parameterinit{4}}
\def\biggfonts{\xviipt}      \def\biggsize{\xviipt\@parameterinit{6}}
\def\Biggfonts{\xxipt}       \def\Biggsize{\xxipt\@parameterinit{8}}
\def\tinyfonts{\vpt}         \def\tinysize{\vpt\@parameterinit{-8}}
\def\HUGEFONTS{\xxvpt}       \def\HUGESIZE{\xxvpt\@parameterinit{10}}

%************************************************************
%*
%*             Document layout
%*
%************************************************************
\message{document layout,}

%************** Page format *********************************

\newtoks\everyoutput \everyoutput={}
\newdimen\depthofpage
\newcount\pagenum \pagenum=0

\newdimen\oddtopmargin  \newdimen\eventopmargin
\newdimen\oddleftmargin \newdimen\evenleftmargin
\newtoks\oddhead        \newtoks\evenhead
\newtoks\oddfoot        \newtoks\evenfoot

\def\topmargin{\afterassignment\@seteventop\oddtopmargin}
\def\leftmargin{\afterassignment\@setevenleft\oddleftmargin}
\def\head{\afterassignment\@setevenhead\oddhead}
\def\foot{\afterassignment\@setevenfoot\oddfoot}

\def\@seteventop{\eventopmargin=\oddtopmargin}
\def\@setevenleft{\evenleftmargin=\oddleftmargin}
\def\@setevenhead{\evenhead=\oddhead}
\def\@setevenfoot{\evenfoot=\oddfoot}

\def\pagenumstyle#1{\@setnumstyle\pagenum{#1}}

\newif\ifdraft
\def\draft{\drafttrue\leftmargin=.5in \overfullrule=5pt }

\def\outputstyle#1{\global\expandafter\let\expandafter
          \@outputstyle\csname#1output\endcsname
     \usename{#1setup}}

\output={\@outputstyle}

\def\normaloutput{\the\everyoutput
     \global\advance\pagenum by\@ne
     \ifodd\pagenum
          \voffset=\oddtopmargin \hoffset=\oddleftmargin
     \else\voffset=\eventopmargin \hoffset=\evenleftmargin
     \fi
     \advance\voffset by-1in  \advance\hoffset by-1in
     \count0=\pagenum
     \expandafter\shipout\pagebox
     \ifnum\outputpenalty>-\@MM\else\dosupereject\fi}

\newdimen\fullhsize
\newbox\leftpage
\newcount\leftpagenum
\newcount\outputpagenum \outputpagenum=0
\let\leftorright=L

\def\twoupoutput{\the\everyoutput
     \global\advance\pagenum by\@ne
     \if L\leftorright
          \global\setbox\leftpage=\leftline{\pagebox}%
          \global\leftpagenum=\pagenum
          \global\let\leftorright=R%
     \else\global\advance\outputpagenum by\@ne
          \ifodd\outputpagenum
               \voffset=\oddtopmargin \hoffset=\oddleftmargin
          \else\voffset=\eventopmargin \hoffset=\evenleftmargin
          \fi
          \advance\voffset by-1in  \advance\hoffset by-1in
          \count0=\leftpagenum \count1=\pagenum
          \shipout\vbox{\hbox to\fullhsize
               {\box\leftpage\hfil\leftline{\pagebox}}}%
          \global\let\leftorright=L%
     \fi
     \ifnum\outputpenalty>-\@MM
     \else\dosupereject
          \if R\leftorright
               \globaldefs=\@ne\head={\hfil}\foot={\hfil}\globaldefs=\z@
               \null\newpage
          \fi
     \fi}

\def\pagebox{\vbox{\makeheadline\pagebody\makefootline}}

\def\makeheadline{%
     \vbox to\z@{\baselinestretch=\@m
          \vskip\topskip\vskip-.708\baselineskip\vskip-\headskip
          \line{\vbox to\ht\strutbox{}%
               \ifodd\pagenum\the\oddhead\else\the\evenhead\fi}%
          \vss}%
     \nointerlineskip}

\def\pagebody{\vbox to\vsize{%
     \boxmaxdepth\maxdepth
     \ifvoid\topins\else\unvbox\topins\fi
     \depthofpage=\dp255
     \unvbox255
     \ifraggedbottom\kern-\depthofpage\vfil\fi
     \ifvoid\footins
     \else\vskip\skip\footins
          \footnoterule
          \unvbox\footins
          \vskip-\footnoteskip
     \fi}}

\def\makefootline{\baselineskip=\footskip
     \line{\ifodd\pagenum\the\oddfoot\else\the\evenfoot\fi}}

%************** Sectioning commands *************************

\newskip\abovechapterskip
\newskip\belowchapterskip
\newskip\abovesectionskip
\newskip\belowsectionskip
\newskip\abovesubsectionskip
\newskip\belowsubsectionskip

\def\chapterstyle#1{\global\expandafter\let\expandafter\@chapterstyle
     \csname#1text\endcsname}
\def\sectionstyle#1{\global\expandafter\let\expandafter\@sectionstyle
     \csname#1text\endcsname}
\def\subsectionstyle#1{\global\expandafter\let\expandafter\@subsectionstyle
     \csname#1text\endcsname}

\def\chapter#1{%
     \ifdim\lastskip=17sp \else\chapterbreak\vskip\abovechapterskip\fi
     \@chapterstyle{\ifblank\chapternumstyle\then
          \else\newchapternum=\next\chapternumformat\ \fi#1}%
     \nobreak\vskip\belowchapterskip\vskip17sp }

\def\section#1{%
     \ifdim\lastskip=17sp \else\sectionbreak\vskip\abovesectionskip\fi
     \@sectionstyle{\ifblank\sectionnumstyle\then
          \else\newsectionnum=\next\sectionnumformat\ \fi#1}%
     \nobreak\vskip\belowsectionskip\vskip17sp }

\def\subsection#1{%
     \ifdim\lastskip=17sp \else\subsectionbreak\vskip\abovesubsectionskip\fi
     \@subsectionstyle{\ifblank\subsectionnumstyle\then
          \else\newsubsectionnum=\next\subsectionnumformat\ \fi#1}%
     \nobreak\vskip\belowsubsectionskip\vskip17sp }

%************** Text formatting commands ********************

\let\TeXunderline=\underline
\let\TeXoverline=\overline
\def\underline#1{\relax\ifmmode\TeXunderline{#1}\else
     $\TeXunderline{\hbox{#1}}$\fi}
\def\overline#1{\relax\ifmmode\TeXoverline{#1}\else
     $\TeXoverline{\hbox{#1}}$\fi}

\def\baselinestretch{\afterassignment\@baselinestretch\count@}
\def\@baselinestretch{\baselineskip=\normalbaselineskip
     \divide\baselineskip by\@m\baselineskip=\count@\baselineskip
     \setbox\strutbox=\hbox{\vrule
          height.708\baselineskip depth.292\baselineskip width\z@}%
     \bigskipamount=\the\baselineskip
          plus.25\baselineskip minus.25\baselineskip
     \medskipamount=.5\baselineskip
          plus.125\baselineskip minus.125\baselineskip
     \smallskipamount=.25\baselineskip
          plus.0625\baselineskip minus.0625\baselineskip}

\def\\{\ifhmode\ifnum\lastpenalty=-\@M\else\hfil\penalty-\@M\fi\fi
     \ignorespaces}
\def\newpage{\vfil\break}

\def\lefttext#1{\par{\@text\leftskip=\z@\rightskip=\centering
     \noindent#1\par}}
\def\righttext#1{\par{\@text\leftskip=\centering\rightskip=\z@
     \noindent#1\par}}
\def\centertext#1{\par{\@text\leftskip=\centering\rightskip=\centering
     \noindent#1\par}}
\def\@text{\parindent=\z@ \parfillskip=\z@ \everypar={}%
     \spaceskip=.3333em \xspaceskip=.5em
     \def\\{\ifhmode\ifnum\lastpenalty=-\@M\else\penalty-\@M\fi\fi
          \ignorespaces}}

\def\beginleft{\par\@text\leftskip=\z@ \rightskip=\centering}
     
\def\beginright{\par\@text\leftskip=\centering\rightskip=\z@ }
     
\def\begincenter{\par\@text\leftskip=\centering\rightskip=\centering}

\def\beginnarrow{\defaultoption[\parindent]\@beginnarrow}
\def\@beginnarrow[#1]{\par\advance\leftskip by#1\advance\rightskip by#1}

\begingroup
\catcode`\[=1 \catcode`\{=11 \gdef\beginignore[\endgroup\bgroup
     \catcode`\e=0 \catcode`\\=12 \catcode`\{=11 \catcode`\f=12 \let\or=\relax
     \let\nd{ignor=\fi \let\}=\egroup
     \iffalse}
\endgroup

\long\def\marginnote#1{\leavevmode
     \edef\@marginsf{\spacefactor=\the\spacefactor\relax}%
     \ifdraft\strut\vadjust{%
          \hbox to\z@{\hskip\hsize\hskip.1in
               \vbox to\z@{\vskip-\dp\strutbox
                    \marginnoteformat
                    \vskip-\ht\strutbox
                    \noindent\strut#1\par
                    \vss}%
               \hss}}%
     \fi
     \@marginsf}

%************** The \bye command ****************************

\newtoks\everybye \everybye={\par\vfil}
\outer\def\bye{\the\everybye
     \footnotecheck
     \prelabelcheck
     \streamcheck
     \supereject
     \TeXend}

%************************************************************
%*
%*             Footnotes
%*
%************************************************************
\message{footnotes,}

\newcount\footnotenum \footnotenum=0
\newskip\footnoteskip
\let\@footnotelist=\empty

\def\footnotenumstyle#1{\@setnumstyle\footnotenum{#1}%
     \useafter\ifx{@footnotenumstyle}\symbols
          \global\let\@footup=\empty
     \else\global\let\@footup=\markup
     \fi}

\def\footnote{\footnotecheck\defaultoption[]\@footnote}
\def\@footnote[#1]{\@footnotemark[#1]\@footnotetext}

\def\footnotemark{\defaultoption[]\@footnotemark}
\def\@footnotemark[#1]{\let\@footsf=\empty
     \ifhmode\edef\@footsf{\spacefactor=\the\spacefactor\relax}\/\fi
     \ifnoarg#1\then
          \global\advance\footnotenum by\@ne
          \@footup{\footnotenumformat}%
          \edef\@@foota{\footnotenum=\the\footnotenum\relax}%
          \expandafter\additemR\expandafter\@footup\expandafter
               {\@@foota\footnotenumformat}\to\@footnotelist
          \global\let\@footnotelist=\@footnotelist
     \else\markup{#1}%
          \additemR\markup{#1}\to\@footnotelist
          \global\let\@footnotelist=\@footnotelist
     \fi
     \@footsf}

\def\footnotetext{%
     \ifx\@footnotelist\empty\err@extrafootnotetext\else\@footnotetext\fi}
\def\@footnotetext{%
     \getitemL\@footnotelist\to\@@foota
     \global\let\@footnotelist=\@footnotelist
     \insert\footins\bgroup
     \footnoteformat
     \splittopskip=\ht\strutbox\splitmaxdepth=\dp\strutbox
     \interlinepenalty=\interfootnotelinepenalty\floatingpenalty=\@MM
     \noindent\llap{\@@foota}\strut
     \bgroup\aftergroup\@footnoteend
     \let\@@scratcha=}
\def\@footnoteend{\strut\par\vskip\footnoteskip\egroup}

\def\footnoterule{\normalfonts
     \kern-.3em \hrule width2in height.04em \kern .26em }

\def\footnotecheck{%
     \ifx\@footnotelist\empty
     \else\err@extrafootnotemark
          \global\let\@footnotelist=\empty
     \fi}

%************************************************************
%*
%*             Labelling macros
%*
%************************************************************
\message{labels,}

\let\@@labeldef=\xdef
\newif\if@labelfile
\newwrite\@labelfile
\let\@prelabellist=\empty

\def\label#1#2{\trim#1\to\@@labarg\edef\@@labtext{#2}%
     \edef\@@labname{lab@\@@labarg}%
     \useafter\ifundefined\@@labname\then\else\@yeslab\fi
     \useafter\@@labeldef\@@labname{#2}%
     \ifstreaming
          \expandafter\toks@\expandafter\expandafter\expandafter
               {\csname\@@labname\endcsname}%
          \immediate\write\streamout{\noexpand\label{\@@labarg}{\the\toks@}}%
     \fi}
\def\@yeslab{%
     \useafter\ifundefined{if\@@labname}\then
          \err@labelredef\@@labarg
     \else\useif{if\@@labname}\then
               \err@labelredef\@@labarg
          \else\global\usename{\@@labname true}%
               \useafter\ifundefined{pre\@@labname}\then
               \else\useafter\ifx{pre\@@labname}\@@labtext
                    \else\err@badlabelmatch\@@labarg
                    \fi
               \fi
               \if@labelfile
               \else\global\@labelfiletrue
                    \immediate\write\sixt@@n{--> Creating file \jobname.lab}%
                    \immediate\openout\@labelfile=\jobname.lab
               \fi
               \immediate\write\@labelfile
                    {\noexpand\prelabel{\@@labarg}{\@@labtext}}%
          \fi
     \fi}

\def\putlab#1{\trim#1\to\@@labarg\edef\@@labname{lab@\@@labarg}%
     \useafter\ifundefined\@@labname\then\@nolab\else\usename\@@labname\fi}
\def\@nolab{%
     \useafter\ifundefined{pre\@@labname}\then
          \undefinedlabelformat
          \err@needlabel\@@labarg
          \useafter\xdef\@@labname{\undefinedlabelformat}%
     \else\usename{pre\@@labname}%
          \useafter\xdef\@@labname{\usename{pre\@@labname}}%
     \fi
     \useafter\newif{if\@@labname}%
     \expandafter\additemR\@@labarg\to\@prelabellist}

\def\prelabel#1{\useafter\gdef{prelab@#1}}

\def\ifundefinedlabel#1\then{%
     \expandafter\ifx\csname lab@#1\endcsname\relax}
\def\useiflab#1\then{\csname iflab@#1\endcsname}

\def\prelabelcheck{{%
     \def\^^\##1{\useiflab{##1}\then\else\err@undefinedlabel{##1}\fi}%
     \@prelabellist}}

%************************************************************
%*
%*             Equation numbering
%*
%************************************************************
\message{equation numbering,}

\newcount\chapternum
\newcount\sectionnum
\newcount\subsectionnum
\newcount\equationnum
\newcount\subequationnum
\newcount\figurenum
\newcount\subfigurenum
\newcount\tablenum
\newcount\subtablenum

\newif\if@subeqncount
\newif\if@subfigcount
\newif\if@subtblcount

\def\newchapternum{\newsectionnum=\z@\@resetnum\chapternum}
\def\newsectionnum{\newsubsectionnum=\z@\@resetnum\sectionnum}
\def\newsubsectionnum{\newequationnum=\z@\newfigurenum=\z@\newtablenum=\z@
     \@resetnum\subsectionnum}
\def\newequationnum{\newsubequationnum=\z@\@resetnum\equationnum}
\def\newsubequationnum{\@resetnum\subequationnum}
\def\newfigurenum{\newsubfigurenum=\z@\@resetnum\figurenum}
\def\newsubfigurenum{\@resetnum\subfigurenum}
\def\newtablenum{\newsubtablenum=\z@\@resetnum\tablenum}
\def\newsubtablenum{\@resetnum\subtablenum}

\def\@resetnum#1{\global\advance#1by1 \edef\next{\the#1\relax}\global#1}

\newchapternum=0

\def\chapternumstyle#1{\@setnumstyle\chapternum{#1}}
\def\sectionnumstyle#1{\@setnumstyle\sectionnum{#1}}
\def\subsectionnumstyle#1{\@setnumstyle\subsectionnum{#1}}
\def\equationnumstyle#1{\@setnumstyle\equationnum{#1}}
\def\subequationnumstyle#1{\@setnumstyle\subequationnum{#1}%
     \ifblank\subequationnumstyle\then\global\@subeqncountfalse\fi
     \ignorespaces}
\def\figurenumstyle#1{\@setnumstyle\figurenum{#1}}
\def\subfigurenumstyle#1{\@setnumstyle\subfigurenum{#1}%
     \ifblank\subfigurenumstyle\then\global\@subfigcountfalse\fi
     \ignorespaces}
\def\tablenumstyle#1{\@setnumstyle\tablenum{#1}}
\def\subtablenumstyle#1{\@setnumstyle\subtablenum{#1}%
     \ifblank\subtablenumstyle\then\global\@subtblcountfalse\fi
     \ignorespaces}

\def\eqnlabel#1{%
     \if@subeqncount
          \newsubequationnum=\next
     \else\newequationnum=\next
          \ifblank\subequationnumstyle\then
          \else\global\@subeqncounttrue
               \newsubequationnum=\@ne
          \fi
     \fi
     \label{#1}{\puteqnformat}(\puteqn{#1})%
     \ifdraft\rlap{\hskip.1in{\tt#1}}\fi}

\let\puteqn=\putlab

\def\equation#1#2{\useafter\gdef{eqn@#1}{#2\eqno\eqnlabel{#1}}}
\def\Equation#1{\useafter\gdef{eqn@#1}}

\def\putequation#1{\useafter\ifundefined{eqn@#1}\then
     \err@undefinedeqn{#1}\else\usename{eqn@#1}\fi}

\def\eqnseriesstyle#1{\gdef\@eqnseriesstyle{#1}}
\def\begineqnseries{\subequationnumstyle{\@eqnseriesstyle}%
     \defaultoption[]\@begineqnseries}
\def\@begineqnseries[#1]{\edef\@@eqnname{#1}}
\def\endeqnseries{\subequationnumstyle{blank}%
     \expandafter\ifnoarg\@@eqnname\then
     \else\label\@@eqnname{\puteqnformat}%
     \fi
     \aftergroup\ignorespaces}

\def\figlabel#1{%
     \if@subfigcount
          \newsubfigurenum=\next
     \else\newfigurenum=\next
          \ifblank\subfigurenumstyle\then
          \else\global\@subfigcounttrue
               \newsubfigurenum=\@ne
          \fi
     \fi
     \label{#1}{\putfigformat}\putfig{#1}%
     {\def\marginnoteformat{\tt}\marginnote{#1}}}

\let\putfig=\putlab

\def\figseriesstyle#1{\gdef\@figseriesstyle{#1}}
\def\beginfigseries{\subfigurenumstyle{\@figseriesstyle}%
     \defaultoption[]\@beginfigseries}
\def\@beginfigseries[#1]{\edef\@@figname{#1}}
\def\endfigseries{\subfigurenumstyle{blank}%
     \expandafter\ifnoarg\@@figname\then
     \else\label\@@figname{\putfigformat}%
     \fi
     \aftergroup\ignorespaces}

\def\tbllabel#1{%
     \if@subtblcount
          \newsubtablenum=\next
     \else\newtablenum=\next
          \ifblank\subtablenumstyle\then
          \else\global\@subtblcounttrue
               \newsubtablenum=\@ne
          \fi
     \fi
     \label{#1}{\puttblformat}\puttbl{#1}%
     {\def\marginnoteformat{\tt}\marginnote{#1}}}

\let\puttbl=\putlab

\def\tblseriesstyle#1{\gdef\@tblseriesstyle{#1}}
\def\begintblseries{\subtablenumstyle{\@tblseriesstyle}%
     \defaultoption[]\@begintblseries}
\def\@begintblseries[#1]{\edef\@@tblname{#1}}
\def\endtblseries{\subtablenumstyle{blank}%
     \expandafter\ifnoarg\@@tblname\then
     \else\label\@@tblname{\puttblformat}%
     \fi
     \aftergroup\ignorespaces}

%************************************************************
%*
%*             Reference numbering
%*
%************************************************************
\message{reference numbering,}

\newcount\referencenum \referencenum=0
\newcount\@@prerefcount \@@prerefcount=0
\newcount\@@thisref
\newcount\@@lastref
\newcount\@@loopref
\newcount\@@refseq
\newdimen\refnumindent
\let\@undefreflist=\empty

\def\referencenumstyle#1{\@setnumstyle\referencenum{#1}}

\def\referencestyle#1{\usename{@ref#1}}

\def\@refsequential{%
     \gdef\@refpredef##1{\global\advance\referencenum by\@ne
          \let\^^\=0\label{##1}{\^^\{\the\referencenum}}%
          \useafter\gdef{ref@\the\referencenum}{{##1}{\undefinedlabelformat}}}%
     \gdef\@reference##1##2{%
          \ifundefinedlabel##1\then
          \else\def\^^\####1{\global\@@thisref=####1\relax}\putlab{##1}%
               \useafter\gdef{ref@\the\@@thisref}{{##1}{##2}}%
          \fi}%
     \gdef\endputreferences{%
          \loop\ifnum\@@loopref<\referencenum
                    \advance\@@loopref by\@ne
                    \expandafter\expandafter\expandafter\@printreference
                         \csname ref@\the\@@loopref\endcsname
          \repeat
          \par}}

\def\@refpreordered{%
     \gdef\@refpredef##1{\global\advance\referencenum by\@ne
          \additemR##1\to\@undefreflist}%
     \gdef\@reference##1##2{%
          \ifundefinedlabel##1\then
          \else\global\advance\@@loopref by\@ne
               {\let\^^\=0\label{##1}{\^^\{\the\@@loopref}}}%
               \@printreference{##1}{##2}%
          \fi}
     \gdef\endputreferences{%
          \def\^^\####1{\useiflab{####1}\then
               \else\reference{####1}{\undefinedlabelformat}\fi}%
          \@undefreflist
          \par}}

\def\beginprereferences{\par
     \def\reference##1##2{\global\advance\referencenum by1\@ne
          \let\^^\=0\label{##1}{\^^\{\the\referencenum}}%
          \useafter\gdef{ref@\the\referencenum}{{##1}{##2}}}}
\def\endprereferences{\global\@@prerefcount=\the\referencenum\par}

\def\beginputreferences{\par
     \refnumindent=\z@\@@loopref=\z@
     \loop\ifnum\@@loopref<\referencenum
               \advance\@@loopref by\@ne
               \setbox\z@=\hbox{\referencenum=\@@loopref
                    \referencenumformat\enskip}%
               \ifdim\wd\z@>\refnumindent\refnumindent=\wd\z@\fi
     \repeat
     \putreferenceformat
     \@@loopref=\z@
     \loop\ifnum\@@loopref<\@@prerefcount
               \advance\@@loopref by\@ne
               \expandafter\expandafter\expandafter\@printreference
                    \csname ref@\the\@@loopref\endcsname
     \repeat
     \let\reference=\@reference}

\def\@printreference#1#2{\ifx#2\undefinedlabelformat\err@undefinedref{#1}\fi
     \noindent\ifdraft\rlap{\hskip\hsize\hskip.1in \tt#1}\fi
     \llap{\referencenum=\@@loopref\referencenumformat\enskip}#2\par}

\def\reference#1#2{{\par\refnumindent=\z@\putreferenceformat\noindent#2\par}}

\def\putref#1{\trim#1\to\@@refarg
     \expandafter\ifnoarg\@@refarg\then
          \toks@={\relax}%
     \else\@@lastref=-\@m\def\@@refsep{}\def\@more{\@nextref}%
          \toks@={\@nextref#1,,}%
     \fi\the\toks@}
\def\@nextref#1,{\trim#1\to\@@refarg
     \expandafter\ifnoarg\@@refarg\then
          \let\@more=\relax
     \else\ifundefinedlabel\@@refarg\then
               \expandafter\@refpredef\expandafter{\@@refarg}%
          \fi
          \def\^^\##1{\global\@@thisref=##1\relax}%
          \global\@@thisref=\m@ne
          \setbox\z@=\hbox{\putlab\@@refarg}%
     \fi
     \advance\@@lastref by\@ne
     \ifnum\@@lastref=\@@thisref\advance\@@refseq by\@ne\else\@@refseq=\@ne\fi
     \ifnum\@@lastref<\z@
     \else\ifnum\@@refseq<\thr@@
               \@@refsep\def\@@refsep{,}%
               \ifnum\@@lastref>\z@
                    \advance\@@lastref by\m@ne
                    {\referencenum=\@@lastref\putrefformat}%
               \else\undefinedlabelformat
               \fi
          \else\def\@@refsep{--}%
          \fi
     \fi
     \@@lastref=\@@thisref
     \@more}

%************************************************************
%*
%*             Job streaming
%*
%************************************************************
\message{streaming,}

\newif\ifstreaming

\def\streamto{\defaultoption[\jobname]\@streamto}
\def\@streamto[#1]{\global\streamingtrue
     \immediate\write\sixt@@n{--> Streaming to #1.str}%
     \newwrite\streamout\immediate\openout\streamout=#1.str }

\def\streamfrom{\defaultoption[\jobname]\@streamfrom}
\def\@streamfrom[#1]{\newread\streamin\openin\streamin=#1.str
     \ifeof\streamin
          \expandafter\err@nostream\expandafter{#1.str}%
     \else\immediate\write\sixt@@n{--> Streaming from #1.str}%
          \let\@@labeldef=\gdef
          \ifstreaming
               \edef\@elc{\endlinechar=\the\endlinechar}%
               \endlinechar=\m@ne
               \loop\read\streamin to\@@scratcha
                    \ifeof\streamin
                         \streamingfalse
                    \else\toks@=\expandafter{\@@scratcha}%
                         \immediate\write\streamout{\the\toks@}%
                    \fi
                    \ifstreaming
               \repeat
               \@elc
               \input #1.str
               \streamingtrue
          \else\input #1.str
          \fi
          \let\@@labeldef=\xdef
     \fi}

\def\streamcheck{\ifstreaming
     \immediate\write\streamout{\pagenum=\the\pagenum}%
     \immediate\write\streamout{\footnotenum=\the\footnotenum}%
     \immediate\write\streamout{\referencenum=\the\referencenum}%
     \immediate\write\streamout{\chapternum=\the\chapternum}%
     \immediate\write\streamout{\sectionnum=\the\sectionnum}%
     \immediate\write\streamout{\subsectionnum=\the\subsectionnum}%
     \immediate\write\streamout{\equationnum=\the\equationnum}%
     \immediate\write\streamout{\subequationnum=\the\subequationnum}%
     \immediate\write\streamout{\figurenum=\the\figurenum}%
     \immediate\write\streamout{\subfigurenum=\the\subfigurenum}%
     \immediate\write\streamout{\tablenum=\the\tablenum}%
     \immediate\write\streamout{\subtablenum=\the\subtablenum}%
     \immediate\closeout\streamout
     \fi}

%************************************************************
%*
%*             Error messages
%*
%************************************************************

\def\err@badtypesize{%
     \errhelp={The limited availability of certain fonts requires^^J%
          that the base type size be 10pt, 12pt, or 14pt.^^J}%
     \errmessage{--> Illegal base type size}}

\def\err@badsizechange{\immediate\write\sixt@@n
     {--> Size change not allowed in math mode, ignored}}

\def\err@sizetoolarge#1{\immediate\write\sixt@@n
     {--> \noexpand#1 too big, substituting HUGE}}

\def\err@sizenotavailable#1{\immediate\write\sixt@@n
     {--> Size not available, \noexpand#1 ignored}}

\def\err@fontnotavailable#1{\immediate\write\sixt@@n
     {--> Font not available, \noexpand#1 ignored}}

\def\err@sltoit{\immediate\write\sixt@@n
     {--> Style \noexpand\sl not available, substituting \noexpand\it}%
     \it}

\def\err@bfstobf{\immediate\write\sixt@@n
     {--> Style \noexpand\bfs not available, substituting \noexpand\bf}%
     \bf}

\def\err@badgroup#1#2{%
     \errhelp={The block you have just tried to close was not the one^^J%
          most recently opened.^^J}%
     \errmessage{--> \noexpand\end{#1} doesn't match \noexpand\begin{#2}}}

\def\err@badcountervalue#1{\immediate\write\sixt@@n
     {--> Counter (#1) out of bounds}}

\def\err@extrafootnotemark{\immediate\write\sixt@@n
     {--> \noexpand\footnotemark command
          has no corresponding \noexpand\footnotetext}}

\def\err@extrafootnotetext{%
     \errhelp{You have given a \noexpand\footnotetext command without first
          specifying^^Ja \noexpand\footnotemark.^^J}%
     \errmessage{--> \noexpand\footnotetext command has no corresponding
          \noexpand\footnotemark}}

\def\err@labelredef#1{\immediate\write\sixt@@n
     {--> Label "#1" redefined}}

\def\err@badlabelmatch#1{\immediate\write\sixt@@n
     {--> Definition of label "#1" doesn't match value in \jobname.lab}}

\def\err@needlabel#1{\immediate\write\sixt@@n
     {--> Label "#1" cited before its definition}}

\def\err@undefinedlabel#1{\immediate\write\sixt@@n
     {--> Label "#1" cited but never defined}}

\def\err@undefinedeqn#1{\immediate\write\sixt@@n
     {--> Equation "#1" not defined}}

\def\err@undefinedref#1{\immediate\write\sixt@@n
     {--> Reference "#1" not defined}}

\def\err@nostream#1{%
     \errhelp={You have tried to input a stream file that doesn't exist.^^J}%
     \errmessage{--> Stream file #1 not found}}

%************************************************************
%*
%*             Initialization
%*
%************************************************************
\message{jyTeX initialization}

\everyjob{\immediate\write16{--> jyTeX version \fmtversion}%
     \edef\@@jobname{\jobname}%
%     \openin0=\inputpath jysupp
%     \ifeof0
%     \else\closein0
%          \immediate\write16{--> Additional macros loaded from jysupp.tex}%
%          \jyinput jysupp
%     \fi
%     \openin0=\inputpath jylocal
%     \ifeof0
%     \else\closein0
%          \immediate\write16{--> Additional macros loaded from jylocal.tex}%
%          \jyinput jylocal
%     \fi
     \edef\jobname{\@@jobname}%
     \settime
     \openin0=\jobname.lab
     \ifeof0
     \else\closein0
          \immediate\write16{--> Getting labels from file \jobname.lab}%
          \input\jobname.lab
     \fi}

%************** Spacing *************************************

\def\fixedskipslist{%
     \^^\{\topskip}%
     \^^\{\splittopskip}%
     \^^\{\maxdepth}%
     \^^\{\skip\topins}%
     \^^\{\skip\footins}%
     \^^\{\headskip}%
     \^^\{\footskip}}

\def\scalingskipslist{%
     \^^\{\p@renwd}%
     \^^\{\delimitershortfall}%
     \^^\{\nulldelimiterspace}%
     \^^\{\scriptspace}%
     \^^\{\jot}%
     \^^\{\normalbaselineskip}%
     \^^\{\normallineskip}%
     \^^\{\normallineskiplimit}%
     \^^\{\baselineskip}%
     \^^\{\lineskip}%
     \^^\{\lineskiplimit}%
     \^^\{\bigskipamount}%
     \^^\{\medskipamount}%
     \^^\{\smallskipamount}%
     \^^\{\parskip}%
     \^^\{\parindent}%
     \^^\{\abovedisplayskip}%
     \^^\{\belowdisplayskip}%
     \^^\{\abovedisplayshortskip}%
     \^^\{\belowdisplayshortskip}%
     \^^\{\abovechapterskip}%
     \^^\{\belowchapterskip}%
     \^^\{\abovesectionskip}%
     \^^\{\belowsectionskip}%
     \^^\{\abovesubsectionskip}%
     \^^\{\belowsubsectionskip}}

%************** Document layout *****************************

\def\twoupsetup{%                                % setup for twoup style
     \topmargin=.75in
     \leftmargin=.5in
     \vsize=6.9in
     \hsize=4.75in
     \fullhsize=10in
     \let\draft=\relax}

\outputstyle{normal}                             % page style

\def\marginnoteformat{\subscriptsize             % paragraphing of margin notes
     \hsize=1in \baselinestretch=1000 \everypar={}%
     \tolerance=5000 \hbadness=5000 \parskip=0pt \parindent=0pt
     \leftskip=0pt \rightskip=0pt \raggedright}

\head={\ifdraft\normalfonts\it\hfil DRAFT\hfil   % format of headline
     \llap{\number\day\ \monthword\month\ \militarytime}\else\hfil\fi}
\foot={\hfil\normalfonts\numstyle\pagenum\hfil}  % format of footline

\normalbaselineskip=12pt                         % usual \baselineskip
\normallineskip=0pt                              % usual \lineskip
\normallineskiplimit=0pt                         % usual \lineskiplimit
\normalbaselines                                 % set \baselineskip

\topskip=.85\baselineskip \splittopskip=\topskip \headskip=2\baselineskip
\footskip=\headskip

\pagenumstyle{arabic}                            % counter style

\parskip=0pt                                     % no skip between paragraphs
\parindent=20pt                                  % usual \parindent

\baselinestretch=1000                            % set \big-, \med-, \smallskip

%************** Sectioning **********************************

\chapterstyle{left}                              % position of heading
\chapternumstyle{blank}                          % counter style
\def\chapterbreak{\newpage}                      % break before heading
\abovechapterskip=0pt                            % space before heading
\belowchapterskip=1.5\baselineskip               % space after heading
     plus.38\baselineskip minus.38\baselineskip
\def\chapternumformat{\numstyle\chapternum.}     % format of heading counter

\sectionstyle{left}                              % position of heading
\sectionnumstyle{blank}                          % counter style
\def\sectionbreak{\vskip0pt plus4\baselineskip\penalty-100
     \vskip0pt plus-4\baselineskip}              % break before heading
\abovesectionskip=1.5\baselineskip               % space before heading
     plus.38\baselineskip minus.38\baselineskip
\belowsectionskip=\the\baselineskip              % space after heading
     plus.25\baselineskip minus.25\baselineskip
\def\sectionnumformat{%                          % format of heading counter
     \ifblank\chapternumstyle\then\else\numstyle\chapternum.\fi
     \numstyle\sectionnum.}

\subsectionstyle{left}                           % position of heading
\subsectionnumstyle{blank}                       % counter style
\def\subsectionbreak{\vskip0pt plus4\baselineskip\penalty-100
     \vskip0pt plus-4\baselineskip}              % break before heading
\abovesubsectionskip=\the\baselineskip           % space before heading
     plus.25\baselineskip minus.25\baselineskip
\belowsubsectionskip=.75\baselineskip            % space after heading
     plus.19\baselineskip minus.19\baselineskip
\def\subsectionnumformat{%                       % format of heading counter
     \ifblank\chapternumstyle\then\else\numstyle\chapternum.\fi
     \ifblank\sectionnumstyle\then\else\numstyle\sectionnum.\fi
     \numstyle\subsectionnum.}

%************** Footnotes ***********************************

\footnotenumstyle{symbols}                       % counter style
\footnoteskip=0pt                                % jyTeX spacing parameter
\def\footnotenumformat{\numstyle\footnotenum}    % \footnotemark format
\def\footnoteformat{\footnotesize                % paragraphing of text
     \everypar={}\parskip=0pt \parfillskip=0pt plus1fil
     \leftskip=1em \rightskip=0pt
     \spaceskip=0pt \xspaceskip=0pt
     \def\\{\ifhmode\ifnum\lastpenalty=-10000
          \else\hfil\penalty-10000 \fi\fi\ignorespaces}}

%************** Labels **************************************

\def\undefinedlabelformat{$\bullet$}             % mark for undefined label

%************** Equation numbering **************************

\equationnumstyle{arabic}                        % counter style
\subequationnumstyle{blank}                      % counter style
\figurenumstyle{arabic}                          % counter style
\subfigurenumstyle{blank}                        % counter style
\tablenumstyle{arabic}                           % counter style
\subtablenumstyle{blank}                         % counter style

\eqnseriesstyle{alphabetic}                      % sub-counter style for series
\figseriesstyle{alphabetic}                      % sub-counter style for series
\tblseriesstyle{alphabetic}                      % sub-counter style for series

\def\puteqnformat{\hbox{%                        % equation number format
     \ifblank\chapternumstyle\then\else\numstyle\chapternum.\fi
     \ifblank\sectionnumstyle\then\else\numstyle\sectionnum.\fi
     \ifblank\subsectionnumstyle\then\else\numstyle\subsectionnum.\fi
     \numstyle\equationnum
     \numstyle\subequationnum}}
\def\putfigformat{\hbox{%                        % figure number format
     \ifblank\chapternumstyle\then\else\numstyle\chapternum.\fi
     \ifblank\sectionnumstyle\then\else\numstyle\sectionnum.\fi
     \ifblank\subsectionnumstyle\then\else\numstyle\subsectionnum.\fi
     \numstyle\figurenum
     \numstyle\subfigurenum}}
\def\puttblformat{\hbox{%                        % table number format
     \ifblank\chapternumstyle\then\else\numstyle\chapternum.\fi
     \ifblank\sectionnumstyle\then\else\numstyle\sectionnum.\fi
     \ifblank\subsectionnumstyle\then\else\numstyle\subsectionnum.\fi
     \numstyle\tablenum
     \numstyle\subtablenum}}

%************** Reference numbering *************************

\referencestyle{sequential}                      % referencing method
\referencenumstyle{arabic}                       % counter style
\def\putrefformat{\numstyle\referencenum}        % format of reference citation
\def\referencenumformat{\numstyle\referencenum.} % format of number in list
\def\putreferenceformat{%                        % paragraphing of list
     \everypar={\hangindent=1em \hangafter=1 }%
     \def\\{\hfil\break\null\hskip-1em \ignorespaces}%
     \leftskip=\refnumindent\parindent=0pt \interlinepenalty=1000 }

%************** Font initialization *************************

\normalsize

%*****************************************************************************

\def\fmtversion{2.6M (June 1992)}

\catcode`\@=12
% ------------------ End of jytex.tex -----------------

%\input jytex.tex   % available from hep-th
\typesize=10pt \magnification=1200 \baselineskip17truept
%\baselineskip25truept
\footnotenumstyle{arabic} \hsize=6truein\vsize=8.5truein
%\input epsf
%\input GJMSren.lab

 %\draft
%\leftmargin=1.25in
%\oddleftmargin=.5in
%\evenleftmargin=1.5in
\sectionnumstyle{blank}
\chapternumstyle{blank}
\chapternum=1
\sectionnum=1
\pagenum=0
%\referencestyle{preordered}
% title style follows

\def\begintitle{\pagenumstyle{blank}\parindent=0pt
\begin{narrow}[0.4in]}
\def\endtitle{\end{narrow}\newpage\pagenumstyle{arabic}}

% exercise style follows

\def\beginexercise{\vskip 20truept\parindent=0pt\begin{narrow}[10
truept]}
\def\endexercise{\vskip 10truept\end{narrow}}

% **************    my jyTeX abbreviations   *****************

\def\eql#1{\eqno\eqnlabel{#1}}
\def\ref{\reference}
\def\peq{\puteqn}
\def\pref{\putref}

\def\mgn{\marginnote}
\def\bex{\begin{exercise}}
\def\eex{\end{exercise}}

% *********************** My definitions ************************

\font\open=msbm10 %scaled\magstep1 % For VAX. Borde p195.

 %scaled\magstep1 % For VAX. Borde p195.
%\font\open=msym10 %scaled\magstep1 % For Arbortxt on PC
%\font\opens=msym8 %scaled\magstep1 % For Arbortxt on PC
\font\goth=eufm10  % For Arbortxt on PC, and VAX. Borde p199

\def\StretchRtArr#1{{\count255=0\loop\relbar\joinrel\advance\count255 by1
\ifnum\count255<#1\repeat\rightarrow}}
\def\StretchLtArr#1{\,{\leftarrow\!\!\count255=0\loop\relbar
\joinrel\advance\count255 by1\ifnum\count255<#1\repeat}}

\def\StretchLRtArr#1{\,{\leftarrow\!\!\count255=0\loop\relbar\joinrel\advance
\count255 by1\ifnum\count255<#1\repeat\rightarrow\,\,}}

\def\mbox#1{{\leavevmode\hbox{#1}}}

\def\hspace#1{{\phantom{\mbox#1}}}
\def\oZ{\mbox{\open\char90}}

\def\gS{\mbox{{\goth\char83}}}

\def\al{\alpha}
 %in jyTeX
 %in jyTeX
 %in jyTeX
 %in jyTeX
 %in jyTeX
 %in jyTeX
 %in jyTeX
 %in jyTeX
 %in jyTeX
 %in jyTeX
 %in jyTeX
 %in jyTeX
 %in jyTeX
% in jyTeX
% in jyTeX
% in jyTeX
\def\bom{{\bmit\omega}}% in jyTeX
% in jyTeX
\def\be{\beta}

\def\de{\delta}
\def\Ga{\Gamma}

\def\ep{\epsilon}

\def\la{\lambda}

\def\om{\omega}
\def\Om{\Omega}

\def\si{\sigma}

\def\ze{\zeta}

\def\caF{{\cal F}}

\def\caQ{{\cal Q}}

\def\caS{{\cal S}}

\def\det{{\rm det\,}}

\def\Real{{\rm Re\,}}

\def\sc{{\rm sc }}

\def\zf{$\zeta$--function}
\def\zfs{$\zeta$--functions}

     % Newline

\def\frac#1/#2{\leavevmode\kern.1em
\raise.5ex\hbox{\the\scriptfont0 #1}\kern-.1em/\kern-.15em
\lower.25ex\hbox{\the\scriptfont0 #2}}
\def\sfrac#1/#2{\leavevmode\kern.1em
\raise.5ex\hbox{\the\scriptscriptfont0 #1}\kern-.1em/\kern-.15em
\lower.25ex\hbox{\the\scriptscriptfont0 #2}}

\def\gtorder{\mathrel{\raise.3ex\hbox{$>$}\mkern-14mu
             \lower0.6ex\hbox{$\sim$}}}
\def\ltorder{\mathrel{\raise.3ex\hbox{$<$}\mkern-14mu
             \lower0.6ex\hbox{$\sim$}}}

\def\semidirprod{\rlap{\ss C}\raise1pt\hbox{$\mkern.75mu\times$}}
\def\for{\lower6pt\hbox{$\Big|$}}
\def\fish{\kern-.25em{\phantom{abcde}\over \phantom{abcde}}\kern-.25em}

 %triple
%dot
 %double
%dot
 %double dot
%for small #1

\def\boxit#1{\vbox{\hrule\hbox{\vrule\kern3pt
        \vbox{\kern3pt#1\kern3pt}\kern3pt\vrule}\hrule}}
\def\dalemb#1#2{{\vbox{\hrule height .#2pt
        \hbox{\vrule width.#2pt height#1pt \kern#1pt \vrule
                width.#2pt} \hrule height.#2pt}}}

        %double stroke
\def\frac#1#2{{{#1}\over{#2}}}
 %lower covariant deriv.
 %upper covariant deriv.
 %lower covariant deriv semicolon.
    %lower ordinary  deriv.
    %lower ordinary  deriv comma.

\def\noin{\noindent}

      %Connection
    %Connection'
\def\comb#1#2{{\left(#1\atop#2\right)}}

\def\nsl{\nabla\!\!\!\! / }

\def\eg{{\it e.g.}}
\def\ie{{\it i.e. }}
\def\cf{{\it cf }}
\def\pa{\partial}

 %gives average <#1>
 %gives thermal average <<#1>>
   %gives bracket <#1|#2>
   %gives comma bracket <#1,#2>
 %gives round bracket (#1,#2)
 %gives round bracket (#1,|#2)
 %gives big bracket <#1|#2>
  %gives
%matrix element <#1|#2|#3>

%gives reduced matrix element
%<#1||#2||#3>

\def\sumdash#1{{\mathop{{\sum}'}_{#1}}}

\def\3j#1#2#3#4#5#6{\left\lgroup\matrix{#1&#2&#3\cr#4&#5&#6\cr}
\right\rgroup}

\def\man{{\cal M}}

\def\m?{\mgn{?}}
% KK's defs

\def\pa{\partial}

\def\beq{\begin{eqnarray}}
\def\eeq{\end{eqnarray}}

%  *******************  Journal refs **********************

\def\cmp#1#2#3{{\it Comm. Math. Phys.} {\bf {#1}} ({#2}) #3}
\def\cqg#1#2#3{{\it Class. Quant. Grav.} {\bf {#1}} ({#2}) #3}

\def\jgp#1#2#3{{\it J. Geom. and Phys.} {\bf {#1}} ({#2}) #3}

\def\jpa#1#2#3{{\it J. Phys.} {\bf A{#1}} ({#2}) #3}

\def\np#1#2#3{{\it Nucl. Phys.} {\bf B{#1}} ({#2}) #3}

\def\pl#1#2#3{{\it Phys. Lett.} {\bf {#1}} ({#2}) #3}

\def\prD#1#2#3{{\it Phys. Rev.} {\bf D{#1}} ({#2}) #3}

\def\am#1#2#3{{\it Acta Mathematica} {\bf {#1}} ({#2}) #3}

\def\dmj#1#2#3{{\it Duke Math. J.} {\bf {#1}} ({#2}) #3}

\def\jpamt#1#2#3{{\it J. Phys.A:Math.Theor.} {\bf{#1}} ({#2}) #3}

\def\ma#1#2#3{{\it Math. Ann.} {\bf {#1}} ({#2}) #3}

\def\plb#1#2#3{{\it Phys. Letts.} {\bf {B#1}} ({#2}) #3}

% *******************   Main text *********************
\begin{title}
\vglue 0.5truein
%\righttext {MUTP/96/23}
%\righttext{hep-th/96}
\vskip15truept
%\leftline{\today}
%\vskip 30truept
\centertext {\Bigfonts \bf  R\'enyi entropy and $C_T$ for higher derivative} \vskip7truept
\vskip10truept\centertext{\Bigfonts \bf free scalars and spinors on even spheres}
 \vskip7truept
\vskip10truept\centertext{\Bigfonts \bf }
 \vskip 20truept
\centertext{J.S.Dowker\footnote{dowkeruk@yahoo.co.uk}} \vskip 7truept \centertext{\it
Theory Group,} \centertext{\it School of Physics and Astronomy,} \centertext{\it The
University of Manchester,} \centertext{\it Manchester, England} \vskip 7truept
\centertext{}

\vskip 7truept \vskip40truept
\begin{narrow}

General expressions for the R\'enyi entropies and central charges for higher derivative free
spinors and scalars on even spheres are obtained using a direct spectral method on a
compact lune division of the sphere. Formulae and numbers are rapidly obtained for any
dimension and order of derivative.

The relation between the conformal anomaly and the hyperbolic free energy is briefly
explored using standard expansions.

A field theoretic derivation of the central charge formula for higher derivative scalars in
any (even) dimension, given by Osborn and Stergiou and by Gliozzi {\it et al}, is thereby
provided. The extension to spinors is made.

Generalised Bernoulli polynomials play an important technical role.

\end{narrow}
\vskip 5truept
%\righttext {August 1996}
\vskip 60truept
%\righttext{Typeset in \jyTeX}
\vfil
\end{title}
\pagenum=0
\newpage

\section{\bf 1. Introduction}

Renyi entropy has recently appeared, as an intermediary, in the computation of the central
charge, and Weyl anomaly, for higher derivative conformal fields, [\pref{BandT}]. The
particular manifold used was the hyperbolic cylinder, on which the propagation operators
factorise. In the present work I will use the periodic spherical $q$--lune for the same
purpose. This is a segment of the $d$--sphere with apex angles, $2\pi/q$. I have
employed this on many previous occasions. The advantage is, partly, its compactness and
the relative ease of higher dimensional calculation.

The R\'enyi entropy for standard fields has been evaluated by Klebanov {\it et al},
[\pref{KPSS}], some time ago. In this, and similar, works the $q$-lune appears as a
`branched sphere' usually interpreted as a covered sphere for which $q=1/n$, with $n$ an
integer. The present evaluations use $q\in\oZ$ which I have always found more
convenient.\footnote{ It is interesting to note that Yankielowicz and Zhou,
[\pref{YandZ}], have recently reached the same conclusion.} The results can be extended
to continuous values of $q$.

\section{\bf 2 R\'enyi entropy}

I repeat a few standard facts for continuity. The R\'enyi entropy, $\gS_n$, is defined by,
$$
   \gS_n={nW(1)-W(1/n)\over1-n}\,,
   \eql{renyi}
  $$
where $W(q)$ here is the effective action on the periodic $q$--lune. In even dimensions, a
universal component of $\gS_n$ is extracted, usually from the coefficient of a logarithmic
term, or a divergent pole, in the effective action. This is the heat--kernel expansion
coefficient, $C_{d/2}$, or, equivalently, the value, $\ze(0)$, of the relevant \zf, up to zero
modes. To a known factor, this is the conformal anomaly.

Considering $n$, or preferably $q$, as a real variable, the expansion of $\gS_n$ about
$n=1$ yields particular information. For example, the first term gives the entanglememt
entropy and the second the central charge, [\pref{Perlmutter}]. This is the route to $C_T$
chosen in [\pref{BandT}].

The technical aim then is to construct the \zf\ of the chosen propagating operator and to
evaluate $\ze(0)$ and its second derivative with respect to $n$, [\pref{Perlmutter}]. I do
this is a direct spectral fashion. The next two sections outline the dynamical situation and
some spectral facts. The basic computation and results are in sections 5 and 6. Section 8
discusses a thermal aspect. The central charge results are in section 10 with the
calculation relegated to Appendix A.

\section{\bf 3 The operators}

All the operators under investigation here take the rather particular Branson spherical
GJMS intertwinor form,
  $$
  \Om_k(d)={\Ga(B+k+1/2)\over\Ga(B-k+1/2)}\,,
  \eql{intertw}
  $$
where $k$ is a real parameter, which will be either an integer or a half--integer. The
(pseudo) operator $B$ depends on the field theory. For scalars $B=B_S= \sqrt{Y_d+1/4}$
where $Y_d$ is the conformally covariant Penrose--Yamabe Laplacian. For Dirac spinors
$B=B_D= (\nsl^{\,2})^{1/2}=|\nsl|$. \footnote{ Gauge forms will be treated at another
time.}

When $k$ is integral or half integral, the form (\peq{intertw}) reduces to a finite product
of operators. To accord with the ordinary field case, $k$ is an integer for integer spin and
a half--integer for half integer spin. These give, respectively, an even or an odd number of
factors. More generally, $k$ has a conformal field theory significance as a scaling
dimension, \cf\ [\pref{Diaz,DandD,KPS2}].

However, since, it turns out, $\ze(0)$ is a rational function of $k$, it is possible (curiously)
to treat all cases by taking, initially, $k$ integral. To get the spinor result one then just
continues in $k$. This saves work. The operator (\peq{intertw}) for integer $k$ reads,
  $$\eqalign{
      \Om_k(d)=&\prod_{j=0}^{k-1} \big(B^2-(j+1/2)^2\big)\cr
      &=\prod_{j=0}^{k-1} \big(B-(j+1/2)\big)\big(B+(j+1/2)\big)\,.\cr
      }
      \eql{op}
  $$

In order to find $\ze(0)$, the spectrum of $\Om_k$, \ie that of $B^2$, is required for the
$q$--lune $d$--manifold. On spherical domains, this spectrum is usually presented in the
form of eigenlevels, depending on one integer, together with their degeneracies which first
have to be determined (no easy task) and then expanded in order to compute the \zf\
usually in terms of the Hurwitz \zf. For the present set of fields this process can be
bypassed, most easily for scalars and spinors. Gauge fields are more involved.

\section{\bf 4. Scalar and spinor spectra on the lune}

I give here a rather abbreviated, heuristic treatment.

The situation is described in [\pref{Dowchem}] for scalars. The standard separation of
variables for the eigenproblem takes on a recursive nature in accordance with the nested
form of the spherical lune metric following from iteration of the geodesic polar coordinate
system. The conical deformation of the lune $d$-manifold can thus be traced ultimately
just to the deformation of the unit circle, S$^1_q$ coordinatised by the polar angle,
$\phi$, which ranges from $0$ to $2\pi/q$.

It was shown in [\pref{Dowchem}] that the spectrum for a charged scalar undergoing a
phase change of $2\pi\de$ on circling S$^1_q$ is the union of two sets of eigenvalues,
$$\eqalign{
&
   \bigg[\big({d-1\over2}+ q|\de|+{\bom}\cdot{\bf n}\big)^2-\al^2\bigg]\cup
   \bigg[\big({d-1\over2}+q-q|\de|+{\bom}\cdot{\bf n}\big)^2-\al^2\bigg]\,,\cr
   }
   \eql{eigsets}
  $$
where $\bom$ is the set of $d$ integers, $(q,\bf1)$, but can be considered, formally, as
real non--negative and referred to as the {\it parameters}. $\bf n$ is a set of $d$
integers, $(n_1,n_2,\ldots,n_d)$ each ranging from 0 to $\infty$. If the propagating
operator is conformal, $\al=1/2$. Degeneracies arise simply from coincidences between
the eigenvalues as presented in (\peq{eigsets}).

In the present work there is no phase change, $\de=0$, and each set of eigenvalues in
(\peq{eigsets}), corresponds, respectively, to Neumann and Dirichlet conditions on the
boundaries of the two `hemilunes', $0\le\phi\le \pi/q$ and $\pi/q\le\phi\le2\pi/q$, that
make up the periodic lune.\footnote{ If $\de=1$, Dirichlet and Neumann are
interchanged.} Also, in this case, one could have real scalars.

For the spinor case, I recall that the Dirac spectrum of $\nsl^2$ on the round sphere
$S^d_1$ is usually expressed in the form of eigenlevels,
  $$
   \la_n(a)=(n+a)^2\,,\quad n=0,1,2,\ldots,\infty\,,\quad a={d\over2}\,,
   \eql{eigs}
  $$
and degeneracies,
  $$
    g_d(n)=2\,2^{[d/2]}\comb {n+d-1}{n}\,.
    \eql{degen}
  $$

I re--express these as follows firstly defining, for convenience,
  $$
  \la({\bf n},a,q)=(a+{\bf n.}\bom)^2\,,\quad {\bom}=(q,1,\ldots,1)\,.
  \eql{lam}
  $$
Then the eigenlevels, (\peq{eigs}), are given by,
  $$
  \la_n(a)=\{\la({\bf n},a,1),n:\sum_i n_i=n\}\,,
  $$
and the degeneracies of the eigenlevels, $\la_n$, arise through coincidences, \ie\
$g_n=\{\sharp \,n_i:\sum_i n_i=n\}$. This gives the binomial coefficient. The 2s are
mostly spin factors.

To find the eigenvalues on the lune, one notes that the sign change in the spinor
components on circumscribing the manifold (\ie around the $S_q^1$) (occasioned by the
rotation of the Zweibeine) can be allowed for by choosing $\de=1/2$ in (\peq{eigsets}),
for scalars. The Dirac eigenvalues then equal,
  $$
  \la({\bf n},a,q)-\al^2,\quad a={d-1+q\over2}\,,
  \eql{Deigs}
  $$
since they give the round values when $q=1$. There is a degeneracy factor of 2 which
reflects the equality of the two sets in (\peq{eigsets}).

These results can be justified  properly by an iterative eigenfunction separation of
variables method detailed, for the round sphere, by Camporesi and Higuchi,
[\pref{CandH}], and on the lune by Apps, [\pref{Apps}] who showed that the two,
identical sets of eigenvalues correspond to the application of the two mixed (local)
conditions at the boundary of a hemilune mentioned above.

The degeneracies on the lune can be calculated, but they become quite complicated,
especially for higher dimensions. Some examples are given by Apps, [\pref{Apps}].
Klebanov {\it et al}, [\pref{KPSS}], give the DIrac expressions in 3 dimensions for the
covering case $q=1/n$. However, there is no need to use them as the forms,
(\peq{eigsets}) and (\peq{Deigs}), can be employed directly, to advantage.

It is worth recalling that the spinor and scalar expressions are related in a well known
thermodynamic fashion that reflects the statistics sign changes. Here, from the specific
forms (\peq{lam}), (\peq{Deigs}), simple rearrangement of the summation gives the
relation between the \zfs,
  $$
  Z_k(s,q,a_S)\bigg|_{spinor}=2\caS\bigg(Z_k(s,q/2,a_N)-Z_k(s,q,a_N)\bigg)\bigg|_{scalar}
  $$
and hence the same relation between derived spectral quantities
  $$
  \caQ_{spinor}(q)=2\caS\big(\caQ_{scalar}(q/2)-Q_{scalar}(q)\big)\,.
  \eql{spsc}
  $$

Invoking this can save effort or act as a check.

\section{\bf 5. Conformal anomaly for product operators}
The spectra outlined in the previous section show that the eigenvalues of the product
$\Om_k$ involve the quantities
  $$
  \la({\bf n},a,q)-\al^2,
  $$
arising from each factor with, explicitly, $\al\equiv\al_j=j+1/2$, $j=0,1,\ldots k-1$. There
are two values of the parameter $a$ to take into account, the Neumann value $a\equiv
a_N=(d-1)/2$ and the Dirichlet one, $a\equiv a_D=a_N+q$ (see (\peq{eigsets})). For
spinors, there are also two values, but they are equal and given in (\peq{Deigs}),
$a\equiv a_S=(d-1+q)/2$. I note that all three values of $a$ coincide when $q=0$. This
means that in this limit (discussed later) the scalar and spinor expressions coincide, as
functions of $k$, up to spin factors.

The object used in the computation of the conformal anomaly is the spectral \zf\ of the
propagation operator, $\Om_k$. I denote this by $Z_k(s)$ with,
  $$\eqalign{
  Z_k(s,q)&=\sum_a Z_k(s,q,a)\cr
  Z_k(s,q,a)&=\caS\,\sum_{j=0}^{k-1}
  \sum_{\bf n=0}^\infty{1\over\big(\la({\bf n},a,q)-\al_j^2\big)^s}\,,\cr
  }
  \eql{zedk}
  $$
where $\caS$ is the bundle dimension. For real scalars $\caS=1$ and for Dirac spinors,
$\caS=2^{d/2}$.

The conformal anomaly for a product, $\Om$, of $k$ second order operators is $k\,Z(0)$
where $Z(s)$ is the co rresponding \zf. This follows, firstly, because the scaling dimension
of the effective action, essentially $\log\det \Om$, is the sum of the scaling dimensions of
the individual factors using linearity of $\log\det$ up to a multiplicative anomaly which
plays no part in scaling. Secondly, $Z(0)$, is the average of the $k$ individual second
order $\ze(0)$s, [\pref{DowGJMS}], equn.(9) using [\pref{Dowcmp}]. Hence the
result.\footnote{This is used in [\pref{Tseytlin,Tseytlin2,Dowconfspins}] for example.}

The computation has already been done in [\pref{DowGJMS}] by the simple expedient of
proceeding to the completely linearised product, as in (\peq{op}). From the structure
(\peq{lam}), only the value of the Barnes \zf\ at $s=0$ is required and this is just a
generalised Bernoulli polynomial, easily evaluated. Barnes' formula is (a residue),

$$
   \ze_d(0,a\,|\,\bom)={(-1)^d\over q\, d!}\,B^{(d)}_d(a\,|\,\bom)\,,
   \quad \bom=(q,1,\ldots,1)\,,
  $$
where the Barnes \zf\ is the continuation of,
  $$\eqalign{ \zeta_d(s, a\mid\bom)
  =&\sum_{{\bf {n}}={\bf 0}}^\infty{1\over(a+{\bf {n.\bom}})^s},\qquad
  \Real\, s>d\,,}
  \eql{barn1}
  $$
so that from (\peq{zedk}), if $d$ is even,
   $$\eqalign{
  kZ_k(0,q,a)&={\caS\over2d!q}\sum_{j=0}^{k-1}\bigg(B_d^{(d)}(a+j+1/2\mid\bom)
  +B_d^{(d)}(a-j-1/2\mid\bom)\bigg)\cr
  &={\caS\over 2\,d!q}\sum_{j=0}^{k-1} \bigg(
  B^{(d)}_d\big(q+d-1-a-j-1/2\mid\bom\big)\cr
  &\hspace{**********}+B^{(d)}_d\big(q+d-1-a+j+1/2\mid\bom\big)\bigg)\,,\cr
  }
  \eql{zezero2}
  $$
using a symmetry property of the Bernoulli polynomials.

Hence we have the equality,
  $$
  Z_k(0,q,a_1)=Z_k(0,q,a_2)\,,
  $$
if
$$
a_1+a_2=d-1+q\,,
$$
which holds for Neumann and Dirichlet conditions and also for spinors (when $a_1=a_2$,
(\peq{Deigs})). Trivially, the sum over $a$ in (\peq{zedk}) gives a factor of 2 in both
cases. Therefore, as a slight notational simplification, I can define
  $$
  C_k(q,a)\equiv(-1)^{2s}\, 2k\,Z_k(0,q,a)\,,
  \eql{zeta0}
  $$
where $s$ is the spin.

The sum in (\peq{zezero2}) can be performed (see Appendix B). This produces the simpler
expression in one higher dimension,
  $$
  C_k(q,a)=(-1)^{2s}{\caS\over\,(d+1)!\,q}\bigg(B^{(d+1)}_{d+1}(a+k
+1/2\mid{\bom,1})-B^{(d+1)}_{d+1}(a-k+1/2\mid{\bom},1)\bigg)\,,
\eql{casimp}
  $$
for the product conformal anomaly.

\section{\bf6. The results}

In accordance with the remarks in section 1, the universal part of the R\'enyi
entropy\footnote{To be clear, this is the coefficient of $\log\ep$ where $1/\ep$ is an IR
cutoff.}, is determined by,

$$
   {\gS}_d(q,k)=-{1\over 1-q}\,\bigg( -q\,C_{k}(q,a)+
   C_{k}(q,a)\big|_{q=1}\bigg)\,,
   \eql{coeff2}
  $$
where the parameter $a$ is $a_N$ (or $a_D$) for scalars, and $a_S$ for spin--half.

The calculation of the Bernoulli polynomials in (\peq{casimp}) is rapid. There is no
practical difficulty in taking the sphere dimension, $d$, as large as desired.

Some polynomial results for a real scalar are,
   $$\eqalign{
   \gS_4(q,k)&=\frac{k\,\left( q+1\right) \,\left( {q}^{2}-10\,{k}^{2}+11\right) }{360}\cr
   \gS_6(q,k)&=-\frac{k\,\left( q+1\right) \,\left( 2\,{q}^{4}-(14\,{k}^{2}
   -23)\,{q}^{2}+42\,{k}^{4}-224\,{k}^{2}+191\right) }{30240}\cr
   \gS_8(q,k)&=\frac{k\,\left( q+1\right)} {1814400}\,\big( 3\,{q}^{6}-(20\,{k}^{2}
   -43)\,{q}^{4}+(42\,{k}^{4}-300\,{k}^{2}+337)\,{q}^{2}\cr
  & \hspace{*****************}-60\,{k}^{6}
  +882\,{k}^{4}-3240\,{k}^{2}+2497\big)\,.
   \cr
  }
  \eql{scalren}
   $$
I remind that $n=1/q$ is the R\'enyi `index'.

If $k$ exceeds $d/2$, the scalar GJMS operators cannot be constructed classically.
However, the above evaluation proceeds without hindrance.

I now list some spinor expressions.
  $$\eqalign{
   \gS_2(q,k)&=-\caS{k(q+1)\over12}\cr
   \gS_4(q,k)&=\caS\frac{k\,\left( q+1\right) \,\left( 7\,{q}^{2}-40\,{k}^{2}
   +47\right) }{1440}\cr
   \gS_6(q,k)&=-\caS\frac{k\,\left( q+1\right) \,\left( 31\,{q}^{4}-196\,{k}^{2}\,{q}^{2}
   +325\,{q}^{2}+336\,{k}^{4}-1876\,{k}^{2}+1669\right) }{241920}\cr
   \gS_8(q,k)&=\caS\!\frac{k\,\left( q+1\right)}{116121600} \,\big( 381\,{q}^{6}
   -(2480\,{k}^{2}-5341)\,{q}^{4}\cr
   &\hspace{**********}+(4704\,{k}^{4}
   -33840\,{k}^{2}+38269)\,{q}^{2}\cr
   &\hspace{************}-3840\,{k}^{6}+58464\,{k}^{4}
   -222000\,{k}^{2}+176509\big)\,. \cr
  }
  \eql{spinren}
$$

As a check, continuing to $k=1/2$ and $q=1$ yields minus the ordinary Dirac conformal
anomaly on the round sphere.

I also list some values for the universal part of the effective action (or free energy),
$\caF_d(q,k)$ (denoted $\caF_q$ by Beccaria and Tseytlin, [\pref{BandT}]), \footnote{
Unfortunately my $q$ is the inverse of that in this reference.}. It equals the coefficient,
$C_k(q,a)$, (\peq{zeta0}). I use both notations.

For scalars,
$$\eqalign{
   \caF_2(q,k)&=\frac{k\,{q}^{2}+2\,{k}^{3}-k}{6\,q}\cr
   \caF_4(q,k)&=-\frac{k\,{q}^{4}+k\left( 1-{k}^{2}\right) \,{q}^{2}
   -6\,{k}^{5}+20\,{k}^{3}-11\,k}{360\,q}\cr
   \caF_6(q,k)&=\frac{1}{30240\,q}\big(2\,k\,{q}^{6}
   +k\left( 3-2\,{k}^{2}\right) \,{q}^{4}\cr
   &\hspace{****}+ 42\,k({k^2}-1)\,(k^2-4)\,{q}^{2}
   +12\,{k}^{7}-126\,{k}^{5}+336\,{k}^{3}-191\,k\big)\,.\cr
  }
  \eql{scaleff}
$$

For spinors,
$$\eqalign{
   \caF_2(q,k)&=\caS\frac{\left( k\,{q}^{2}-4\,{k}^{3}+2\,k\right) }{6\,q}\cr
   \caF_4(q,k)&=-\caS\frac{\left( 7\,k\,{q}^{4}+40k\left( 1-
   {k}^{2}\right) \,{q}^{2}+48\,{k}^{5}-160\,{k}^{3}
   +88\,k\right)}{1440\,q}\cr
   \caF_6(q,k)&=\frac \caS{241920\,q}\big( 31\,k\,{q}^{6}+98k( 3-2k^2) \,{q}^{4}+\cr
   &\hspace{****}336k(k^2-1)(k^4-4)\,{q}^{2}-192\,{k}^{7}+2016\,{k}^{5}-5376\,{k}^{3}
   +3056\,k\big)\,.\cr
  }
  \eql{spineff}
$$

The relation (\peq{spsc}) could be checked at this point.

For scalars, there is no term proportional to $q$ at the subcritical, physical values
$k=1,2,\ldots,d/2-1$. This is a consequence of the conformal invariance. The same is not
true for spinors at the Dirac physical values, $k=1/2,\ldots$.

\section{\bf7. Comparison}

We can make contact with previous evaluations of some of these quantities. Beccaria and
Tseytlin, [\pref{BandT}], compute the $|\nsl|^3$ entropy (corresponding to $k=3/2$ in
(\peq{spinren})) in four and six dimensions.\footnote{ The expressions in this reference
differ in sign from the ones here because they are the universal coefficients of
$\log(1/\ep)$.} They employ the popular hyperbolic cylinder method advocated by Casini
and Huerta, [\pref{CaandH}], in their evaluation of R\'enyi entropy for standard fields on
spheres. The propagating operators again factorise on the cylinder. For standard fields,
$k=1$, the expressions (\peq{scalren}) agree with those in [\pref{CaandH}] and
[\pref{Dowren}]. An alternative derivation is in [\pref{Dowchem}].

A comparison of the free energy expressions here and in [\pref{BandT}] shows that they
differ by a term proportional to $1/q$ which does not affect the R\'enyi entropy, As $q\to
0$, which is a low temperature limit, the coefficient of $1/q$,  $\sim$ the inverse
temperature, is related to the vacuum energy on the odd $(d-1)$--sphere. This numerical
statement is made a little more analytical in the next section.

\section{\bf 8. $q\to 0$ limit. Thermal interpretation and vacuum energy}

The value $q\,C_k(q,a)$ at $q=0$ is easily obtained from (\peq{zeta0}) with
(\peq{zezero2}).

From their definition, the generalised Bernoulli polynomials $B^{(n)}_\nu\big(x(q)\mid
q,{\bf1}\big)$ reduce, at $q=0$ to,
  $$
  B^{(n)}_\nu\big(x(0)\mid 0,{\bf1}\big)=B^{(n-1)}_\nu\big(x(0)\big)\,,
  $$
where $B^{(n-1)}_\nu(x)\equiv B^{(n-1)}_\nu\big(x\mid {\bf1})$, all the parameters
equaling unity. (This is usually referred to as {\it the} generalised Bernoulli polynomial.)
They are easily calculated.

Hence, making use of the simplification derived in Appendix B,
    $$
    q\,\caF_d\big(q,k\big)\big|_{q=0}=(-1)^{2s}{\caS\over(d+1)!}
    \bigg(B_{d+1}^{(d)}(a_0+k+1/2)
  -B_{d+1}^{(d)}(a_0-k+1/2)\bigg)\,,
    $$
where $a_0= (d-1)/2$ for both scalars and spinors.

I now recall the expression for the vacuum energy for GJMS operators on the Einstein
cylinder $R\times$ S$^{d-1}$, with $d$ even, obtained and computed in
[\pref{dowqretspin}].

In particular for scalars,
$$
  \eqalign{
  E_0(d,k)&=-{1\over2(d+1)!}\bigg(B^{(d)}_{d+1}\big(d/2+k\big)-
  \,B^{(d)}_{d+1}\big(d/2-k\big)\bigg)\cr
  &=-{1\over(d+1)!}
  \,B^{(d)}_{d+1}\big(d/2+k\big)\,,\cr
  }
  $$
One sees therefore that the coefficient of $1/q$ in the free energy, $\caF_d(q,k)$, is
minus twice the vacuum energy, $E_0$, on the Einstein cylinder.

In order to give this result a hyperbolic thermal interpretation it is best to restate it as the
coefficient of $2\pi/q$ in (the universal part of) the free energy is $-\pi\,E_0$ which can be
taken as a (regularised) vacuum energy on the open Einstein cylinder R$\times$H$^{d-1}$
since the volume of S$^d$ is $\pi$ times the regularised volume of H$^d$, up to an
alternating sign. This means that the free energy density on S$^1_q\times $H$^{d-1}$
tends to $-\be |\eta_0|$ as $\be=2\pi/q\to\infty$ where $\eta_0$ is the vacuum energy
density on S$^{d-1}$ which also alternates in sign.\footnote{ The signs all stem from
dimensions and the fact that a hyperboloid is a sphere of imaginary radius.} By contrast, a
conventional direct field theory calculation of the local energy density on the open Einstein
cylinder actually yields zero.

Spinors can be discussed likewise.

\section{\bf 9. Entanglement entropy}

The (universal part of) the entanglement entropy arises as the $q\to1$ value of the
R\'enyi entropy and is readily found. For standard fields it equals (minus) the conformal
anomaly. This is a consequence of the stationarlty of the heat--kernel coefficient at $q=1$
as proved in [\pref{Dowhyp}]. A similar statement holds in the case of product operators,
as I now show in a very similar fashion.

One needs to calculate the derivative $\pa C_k(q,a)/\pa q$ using (\peq{casimp}). The
necessary formula, due to Barnes,  was used in [\pref{Dowhyp,Dowpiston}]. I repeat it
here,
   $$
    \pa_q\,{1\over q}\,B^{(d)}_{d}(x|\,q,{\bf1})\bigg|_{q=1}=
    -\,B^{(d+1)}_{d}(x+1|{\bf1})\,.
    \eql{diff}
  $$

This gives ( I detail scalars only),
  $$\eqalign{
 {\pa C_k(q,a)\over\pa q}\bigg|_{q=1}&={1\over\,(d+1)!}
 \bigg(-B^{(d+2)}_{d+1}(d/2+k
+1\mid{\bf1})+B^{(d+2)}_{d+1}\big(d/2-k+1\mid{\bf1}\big)\bigg)\cr
&=-2{1\over\,(d+1)!}
 B^{(d+2)}_{d+1}\big(d/2+k+1\mid{\bf1}\big)\cr
}
\eql{casimp2}
  $$
after using a symmetry of the Bernoulli polynomials ( $d$ is even).

The simple product structure,
  $$
    B^{(d+2)}_{d+1}(x\big|\,{\bf 1})=(x-1)(x-2)\ldots(x-d-1)\,,
    \eql{prods}
  $$
then trivially shows that the derivative is zero if $k$ is an integer in the range $-d/2\le
k\le d/2$. The spinor result is that $k$ is a half--integer with $(1-d)/2\le k\le (1+d/)2$.
The positive values are the relevant ones.

\section{\bf 10. Derivatives and central charge}

It is easy to compute, or read off, the derivatives of the R\'enyi entropy at $q=1$. As in
[\pref{dowrenexp}], giving the expressions as polynomials in $q$ is convenient for this.
Unfortunately from the explicit forms, one has to proceed dimension by dimension but
further work can produce a more compact result whose form can be compared with existing
ones. The details appear in Appendix A where it is shown that the required derivative (I
exhibit the free energy) is given by,
$$
  \caF''_d(k)\equiv{\pa^2 \caF_d(q,k)\over\pa n^2}\bigg|_{n=1}=(-1)^{d/2+k}
  {4k\over d+2}{(d/2+k)!(d/2-k)!\over(d+1)!}\,,
  \eql{cad}
  $$
for scalars.

From its derivation, this formula works only for $k$ an integer in the range $-d/2\le k\le
d/2$. At these values it agrees with the polynomial (in $k$) expressions obtained, for each
dimension, from the results in (\peq{scaleff}). For example,
$$\eqalign{
   \caF''_2(k)&={k\over3}\cr
   \caF''_4(k)&=\frac{5\,{k}^{3}-8\,k}{90}\cr
   \caF''_6(k)&=\frac{7\,{k}^{5}-49\,{k}^{3}+54\,k}{2520}\,,\cr
  }
  \eql{scaleffd}
$$
which can therefore be taken as some sort of continuation of the general, but restricted,
formula, (\peq{cad}).\footnote{ The values supplied by (\peq{cad}) could be used to fix
the form of an {\it unknown} polynomial.}

This formula also leads to an expression for the scalar central charge using the relations
given by Perlmutter, [\pref{Perlmutter}]. I find,
$$
C_T(d,k)=(-1)^{k+1}{4k {(d/2+k)!\,(d/2-k)!}\over (d+2){\left( d-1\right) \,
{\left( d/2-1\right) !}^{2}}}\,.
\eql{ceet}
  $$

Computation reveals agreement with values occurring in [\pref{BandT}] \footnote{ My use
of a UV log cutoff accounts for any sign differences.} which reference also details earlier
calculations.

I list a few numerical values from (\peq{ceet})
  $$\eqalign{
C_T(6,k)&=\frac{6}{5},-6,54\cr
C_T(8,k)&=\frac{8}{7},-\frac{32}{7},24,-256\cr
C_T(10,k)&=\frac{10}{9},-\frac{35}{9},\frac{140}{9},-\frac{280}{3},\frac{3500}{3}\,.\cr
  }
  \eql{ceetscv}
  $$

Osborn and Stergiou, [\pref{OandS}], address higher derivative fields (restricted to
$k\le3$) using CFT techniques. My evaluation furnishes a self--contained proof of their
conjecture for $C_T$ for any $k$ which employs only simple spectral methods.

Based on work by Guerrieri, Petkou and Wen, [\pref{GPW}], a general demonstration was
outlined in [\pref{OandS}] which imported an expression for the CFT four--point function
derived elsewhere with some work. A CFT proof was also provided by Gliozzi {\it et al},
[\pref{GGPW}].

I find the corresponding spinor general expression to be (briefly mentioned in Appendix A),
  $$
  C_T(d,k)=(-1)^{l}\caS
  {\big(8l(l+1)-(d+2)(d-1)\big)(d/2+l)!(d/2-l-1)!\over(d+2)(d-1)\big((d/2-1)!\big)^2}\,.
  \eql{ctspin}
  $$

For uniformity I have used the half--integer (for spinors) $k,=l+1/2,$ as the argument.

Any numerical value is quickly generated from the previous explicit results, \eg\
(\peq{spineff}), and the Perlmutter factor, or from (\peq{ctspin}). As examples,
  $$\eqalign{
C_T(4,k)&=2\,S,-\frac{2\,S}{3}\cr
C_T(6,k)&=3\,S,-\frac{18\,S}{5},-6\cr
C_T(8,k)&=4\,S,-\frac{36\,S}{7},\frac{44\,S}{7},52\,S\cr
C_T(10,k)&=5\,S,-\frac{115\,S}{18},\frac{175\,S}{18},-\frac{70\,S}{9},
-\frac{910\,S}{3}\cr
C_T(12,k)&=6\,S,-\frac{414\,S}{55},\frac{636\,S}{55},-\frac{1044\,S}{55},
-\frac{108\,S}{11},1548\,S\,,
  }
  \eql{ceetspv}
  $$
where $k$ runs from $1/2$ to $(d-1)/2$, $2k$ being the derivative power of the spinor.

In the notation of [\pref{BandT}], $C_T(d,k)=C_{T,d}\big(\chi^{(2k)}\big)$ where $\chi$
is a generic field (scalar or spinor). In particular, one sees that
$C_{T,6}\big(\psi^{(3)}\big)=-18\,\caS/5$, the value preferred in [\pref{BandT}],
($\caS=n_F$). I am not aware of any significance to the higher values.
\section{\bf11. Comments}

The present calculation is a compact version of that in [\pref{BandT}], but one that allows
more general expressions to be derived efficiently and numbers to be found rapidly.

The formula for the central charge relies on the relation with the derivatives of the free
energy derived by Perlmutter, [\pref{Perlmutter}], in a non--CFT way.

The precise nature of the continuation provided by the polynomials, \eg\ (\peq{scaleffd}),
in the number of quadratic factors, $k$, remains to be elucidated in the light of the
general expression, ({\peq{intertw}), which implies, {\it inter alia},
$\Om_{-k}=\Om_k^{-1}$.

Gauge fields present extra technical difficulties connected with the ghost sums.

The thermal aspects in section 8 need clarification.

The termination of the heat--kernel expansion on odd $d-1$--spheres, and pseudospheres,
for conformality in $d$, has been known and used for a very long time as have its
implications for the finite temperature field theory. In particular for pseudospheres the
high temperature expansion, of, say, the free energy is both finite and {\it exact},
[\pref{CandD}]. For spheres there is a non--local exponentially small component reflecting
the existence of closed geodesics, responsible for the Casimir energy.

In the present calculation, the pseudosphere exactness accounts for the finite nature of
the free energy polynomials in $q$, (\peq{scaleff}), by noting the relation of these with
the hyperbolic expressions and that these last are derived by the same method given
originally in [\pref{DandK}], where general expansions can be found. Some further details
are given in Appendix C.

The calculation here via the conically deformed $d$--sphere, S$^d_q$, shows that the
heat--kernel coefficients on the round $(d-1)$--sphere, S$^{d-1}_1$, (they are the same
as on the hyperboloid, H$^{d-1}$, up to signs) can be obtained from the conformal
anomaly on S$^d_q$. They are, of course, well known.

The corresponding calculation for odd spheres will require the calculation of the functional
determinants of the operator, $\Om_k$.

\section{\bf Appendix A}

I here calculate the second derivative of the conformal anomaly (equivalently the `free
energy'), $\pa^2 C_k(q,a)/\pa n^2$ at $n=1$. It is better to use $q=1/n$ so for scalar
fields, \ie\ $a=(d-1)/2$,  I require, from (\peq{casimp}),
  $$
   q^2{\pa\over\pa q}\,q^2\,{\pa\over\pa q}\,{1\over q}B^{(d+1)}_{d+1}(b\mid
   q,{\bf1})=  - q^2{\pa\over\pa q}B^{(d+2)}_{d+1}(b+q\mid q,q,{\bf1})\,,
   \eql{2deriv2}
  $$
at $q=1$ with the integer $b=d/2\pm k$.

The algebra is performed in [\pref{Dowpiston}] and I find,
  $$\eqalign{
  -{\pa\over\pa q}&B^{(d+2)}_{d+1}(b+q\mid q,q,{\bf1})\bigg|_{q=1}=\cr
  &2B^{(d+3)}_{d+1}(b+2\mid {\bf1})-
  2B^{(d+2)}_{d+1}(b+1\mid {\bf1})
  -(d+1)\,B^{(d+2)}_{d}(b+1\mid {\bf1})\,,
  }
  \eql{2deriv3}
  $$
where the second term is immediately zero in light of the product (\peq{prods}) and the
allowed range of $k$.

The expression can be reduced by using the recursion,
$$
  B^{(d+3)}_{d+1}(b+2)-B^{(d+3)}_{d+1}(b+1)=(d+1)\,B^{(d+2)}_{d}(b+1)\,,
  \eql{genB}
  $$
to obtain for (\peq{2deriv3}),
$$
B^{(d+3)}_{d+1}(b+2)+
  B^{(d+3)}_{d+1}(b+1)\,.
  \eql{2deriv4}
  $$
I have dropped the reference to the parameters, which are now all unity so that use can
be made of existing explicit expressions, [\pref{Norlund}], for these Bernoulli polynomials,
\eg
  $$
  B^{(d+3)}_{d+1}(x)={1\over d+2}\sum_{i=1}^{d+2} (x-1)(x-2)\,,
  \ldots\widehat{(x-i)}\ldots(x-d-2)\,,
  $$
omitting the $i$th factor.

There is always a vanishing factor for the polynomials in (\peq{2deriv4}), if $k$ is in the
allowed range. Substitution and combination produce the final answer for the second
derivative of the scalar free--energy,
  $$
  {\pa^2 C_k(q,a)\over\pa n^2}\bigg|_{n=1}=4(-1)^{d/2+k}
  {k(d/2+k)!(d/2-k)!\over(d+2)\,(d+1)!}\,.
  $$

The spinor calculation is a little more complicated because the parameter $a_S$  depends
on $q$ from the start.

In place of (\peq{2deriv4}) there results,
  $$
-{1\over2}B^{(d+3)}_{d+1}(b+3)-3B^{(d+3)}_{d+1}(b+2)-
{1\over2}\,B^{(d+3)}_{d+1}(b+1)\,,
  \eql{2deriv5}
  $$
and for the second derivative of the spinor free energy, ($l+1 $ is the order of the Dirac
operator), simple algebra reveals,
  $$
  {\pa^2 C_k(q,a)\over\pa n^2}\bigg|_{n=1}=(-1)^{d/2+l}\caS
  {\big(8l(l+1)-(d+2)(d-1)\big)\big(d/2+l\big)!\big(d/2-l-1\big)!\over(d+2)\,(d+1)!}\,.
  $$
This yields the formula for the spinor central charge given in section 10.

\section{\bf Appendix B}

The summation over $j$ in (\peq{zezero2}) is here performed. This has handy formal
consequences. The Barnes \zf\ has the Hankel contour representation,
$$\eqalign{ \zeta_d(s,a\mid{\bom})=&{i\Gamma(1-s)\over2\pi}\int_L
  d\tau {\exp(-a\tau)
  (-\tau)^{s-1}\over\prod_{i=1}^d\big(1-\exp(-\om_i\tau)\big)}\,.\cr
  }
  \eql{barn}
  $$

Set $a\to a\pm(j+1/2)$ and perform the geometric sum over $j$. Simple algebra yields,
 $$\eqalign{
&\sum_{j=0}^{k-1}\big(\zeta_d(s,a+j+1/2\mid{\bom})+\zeta_d(s,a-j-1/2\mid{\bom})\big)\cr
&=\ze_{d+1}(s,a+1/2-k\mid \bom,1)-\ze_{d+1}(s,a+1/2+k\mid\bom,1)\,.
}
 $$

The $s=0$ values provide the specific simplification,
 $$\eqalign{
&\sum_{j=0}^{k-1}\big(B^{(d)}_d(a+j
+1/2\mid{\bom})+B^{(d)}_d(a-j-1/2\mid{\bom})\big)\cr
&={1\over\,d+1}\big(B^{(d+1)}_{d+1}(a+k
+1/2\mid{\bom,1})-B^{(d+1)}_{d+1}(a-k+1/2\mid{\bom},1)\big)\,.\cr
}
\eql{bsimp}
 $$

Other values for $s$ give  other relations.

\section{\bf Appendix C}

It is interesting to follow through the relation between the conformal anomaly on the
$q$--deformed sphere and the {\it conventional} high temperature expansion of the free
energy on the open Einstein cylinder. For simplicity, I consider only standard scalars so
that many formulae will have appeared before \eg\ in [\pref{Dowhyp}]. Some of the
general polynomial structure behind the specific results like (\peq{scaleff}) will thereby
appear.

 The general high temperature expansions in an  ultrastatic space--time,
$R\times\man$, were early derived in [\pref{DandK}] in terms of the heat--kernel
coefficients, $c_m$, on $\man$, and extended to any dimension in [\pref{dowvacenhd}].
The relevant part of the series for the {\it finite temperature} free energy is ($\man$ is
supposed closed, and of odd dimension, $d-1$),
  $$\eqalign{
\be F&\sim-{1\over \pi^{1/2}}\sumdash{m=0,1,\ldots} {c_m(d)\over 2^{2m-d+1}}
\,\ze_R(d-2m)\,\Ga(d/2-m)\,\be^{2m-d+1}-{1\over2}\ze_\man'(0)\cr
&\sim{2\pi\over\pi^{1/2}}\sumdash{m=0,1,\ldots}\!\!\!\! c_m(d)
\,(-1)^{d/2-m}{2^{d-2m-3}B_{d-2m}\over (d-2m)!}\,\Ga(d/2-m)\,q^{d-2m-1}
-{1\over 2}\ze_\man'(0)\cr
}
\eql{betaF}
  $$
where the dash on the summation indicates that the term $m=d/2$ is to be omitted and
where $q\equiv 2\pi/\be$.

The final term is the formal effect of the zero mode on the thermal circle. In the hyperbolic
calculations this mode is dropped as a renormalisation.

Apply (\peq{betaF}) to $\man=$ S$^{d-1}$. The transition to the pseodosphere,
H$^{d-1}$, is made shortly.

The heat--kernel coefficients, $c_m$, have been given, in the form I need, in
[\pref{ChandD}]. They are zero from at least $m=d/2$ onwards and
  $$\eqalign{
  c_m(d)&={2\,\Ga((d-1)/2-m)\over(d-2-2m)!\,(2m)!}\,B^{(d-1)}_{2m}
  \big(d/2)\cr
&=\sqrt\pi{2^{2m+3-d}\over(d/2-1-m)!\,(2m)!}\,B^{(d-1)}_{2m}
  \big(d/2)\,,\quad m=0,1,\ldots\cr
  }
  \eql{ceem}
  $$

Moreover, the product form of the Bernoulli polynomial $B^{(d-1)}_{d-2}$ shows that the
coefficients actually vanish from $m=d/2-1$ onwards. All this is well known.

Substituting (\peq{ceem}) into (\peq{betaF}),
$$\eqalign{
\be F
&=2\pi\sum_{m=0}^{d/2-2}
\,(-1)^{d/2-m}{B_{d-2m}B^{(d-1)}_{2m}
  \big(d/2)\over (2m)!(d-2m)!}\,q^{d-2m-1}-{1\over 2}\ze_\man'(0)\,.\cr
}
\eql{betaF2}
  $$

I now turn to the conformal anomaly which is, from (\peq{zezero2}),
$$\eqalign{
  C_1(q,a)&={1\over d!q}\bigg(B_d^{(d)}(d/2\mid\bom)
  +B_d^{(d)}(d/2-1\mid\bom)\bigg)\,.\cr
  }
  \eql{ca1}
  $$

To relate this to (\peq{betaF2 }), I invoke the expansion ([\pref{Norlund}] p.165, (23)),
 $$\eqalign{
  B^{(d)}_d(x\mid\bom)&=\sum_{\mu=0}^d\comb{d}
  {\mu}q^{d-\mu}\,B_{d-\mu}\,B^{(d-1)}_{\mu}(x)\cr
  &=-{d\over2}qB^{(d-1)}_{d-1}(x)+\sum_{m=0}^{d/2}\comb{d}
  {2m}q^{d-2m}\,B_{d-2m}\,B^{(d-1)}_{2m}(x)\cr
  }
 $$
which gives the structure of the $q$--polynomials.

Addition of the two terms in (\peq{ca1}) and use of a symmetry of the Bernoulli
polynomials produces
  $$\eqalign{
  {1\over qd!}\bigg(B^{(d)}_d(d/2\mid\bom)&+B^{(d)}_d(d/2-1\mid\bom)\bigg)
  =2\sum_{m=0}^{d/2}
  q^{d-2m-1}{B_{d-2m}\,B^{(d-1)}_{2m}(d/2)\over(2m)!(d-2m)!}\cr
&=2\sum_{m=0}^{d/2-2}q^{d-2m-1}{B_{d-2m}\,B^{(d-1)}_{2m}(d/2)\over
(2m!)(d-2m)!}+{2\over d!\,q}B^{(d-1)}_d(d/2)\,.\cr
  }
  \eql{ca3}
 $$

There are various ways of organising  the heat--kernel coefficients on the $q$--deformed
sphere as functions of $q$, some more expressive than others, but the above is sufficient
for my purpose, at the moment.

To finalise the equivalence, it is necessary to convert the {\it sphere} free energy, $F$, to
the {\it pseudosphere} value. To do this, I first pass to densities by dividing the series
extensive part of (\peq{betaF2}) by the volume of the $(d-1)$--sphere, then changing the
signs of the resulting (local) $c_m$ coefficients by $(-1)^m$ to give those on the
pseudosphere and finally multiplying by the (regularised) volume of $H^{d-1}$ to give the
`global' pseudosphere values. The ratio of volumes removes the $(-1)^{d/2}\,\pi$,
leaving a factor of 2. Comparison of the result for $\beta F$ with the conformal anomaly,
(\peq{ca3}), then shows that the series terms in each are identical.

The final term in $\be F$, (\peq{betaF2}) is global and non--extensive.\footnote{ It does
not contribute to the internal energy but adds to the entropy. It has been termed a
`topological entropy' by Asorey {\it et al}, [\pref{ABCDS,ABAS}] }  Like the Casimir
energy, it is occasioned by the existence of closed (finite) geodesics on the sphere and so
would not carry over to the non--compact hyperboloid.

As mentioned earlier, because of the absence of exponentially small corrections to the
asymptotic heat--kernel expansion, and also of the termination of this, the high
temperature series for the hyperbolic case is actually exact.

The final term in the conformal anomaly, (\peq{ca3}), does not appear in $\be F$. This
term is minus twice the Casimir energy on the closed Einstein cylinder, [\pref{ChandD}],
as has been shown more generally in section 8.

 \vglue 20truept

 \noin{\bf References.} \vskip5truept
\begin{putreferences}
  \ref{BandT}{Beccaria,M. and Tseytlin,A.A. {\it $C_T$ for higher derivative conformal
  fields and anomalies of (1,0) superconformal 6d theories}, ArXiv:1705.00305.}
  \ref{ABCDS}{Asorey,M, Beneventano, Calvero-Pel\'aes, I, C.G., D'Ascanio, D. and Santangelo,
  E.M.{\it Topological entropy and renormalization group flow in 3--dimensional spherical
  spaces}, ArXiv:1406.6602.}
  \ref{ABAS}{Asorey,M, Beneventano, C.G., D'Ascanio, D. and Santangelo, E.M.
    {\it Thermodynmics of conformal fields in topologically non--trivial space--time
    backgrounds}, ArXiv:1212.6220.}
  \ref{GPW}{Guerrieri, A.L., Petkou, A. C. and Wen, C. {\it The free $\si$CFTs},
  ArXiv:1604.07310.}
  \ref{GGPW}{Gliozzi,F., Guerrieri, A.L., Petkou, A.C. and Wen,C.
   {\it The analytic structure of conformal blocks and the
   generalized Wilson--Fisher fixed points}, {\it JHEP }1704 (2017) 056, ArXiv:1702.03938.}
  \ref{YandZ}{Yankielowicz, S. and Zhou,Y. {\it Supersymmetric R\'enyi Entropy and
  Anomalies in Six--Dimensional (1,0) Superconformal Theories}, ArXiv:1702.03518.}
  \ref{OandS}{Osborn.H. and Stergiou, A. {\it $C_T$ for Non--unitary CFTs in higher dimensions},
  {\it JHEP} {\bf06} (2016) 079, ArXiv:1603.07307.}
  \ref{CandD}{Candelas,P. and Dowker,J.S. \prD{19}{1979}{2902}.}
  \ref{Perlmutter}{Perlmutter,E. {\it A universal feature of CFT R\'enyi entropy}
  {\it JHEP} {\bf03} (2014) 117. ArXiv:1308.1083.}
   \ref{Norlund}{N\"orlund,N.E. {\it M\'emoire sur les polynomes de Bernoulli}, \am{43}{1922}{121}.}
   \ref{dowqretspin}{Dowker,J.S. {\it Revivals and Casimir energy for a free Maxwell field
  (spin-1 singleton) on $R\times S^d$ for odd $d$}, ArXiv:1605.01633.}
   \ref{Dowpiston}{Dowker,J.S. {\it Spherical Casimir pistons}, \cqg{28}{2011}{155018},
   ArXiv:1102.1946.}
  \ref{dowvacenhd}{Dowker,J.S. {\it Finite temperature and vacuum effects in
  higher dimensions}. \cqg{1}{1984}{359}.}
  \ref{Dowchem}{Dowker,J.S. {\it Charged R\'enyi entropy for free scalar fields}, \jpa{50}
  {2017}{165401}, ArXiv:1512.01135.}
  \ref{Dowconfspins}{Dowker,J.S. {\it Effective action of conformal spins on spheres
  with multiplicative and conformal anomalies}, \jpa{48}{2015}{225402}, ArXiv:1501.04881.}
  \ref{Dowhyp}{Dowker,J.S. {\it Hyperspherical entanglement entropy},
  \jpa{43}{2010}{445402}, ArXiv:1007.3865.}
   \ref{DandK}{Dowker,J.S. and Kennedy,G. \jpa{11}{1978}{895}.}
  \ref{dowrenexp}{Dowker,J.S.{\it Expansion of R\'enyi entropy for free scalar fields},
   ArXiv:1412.0549.}
     \ref{CaandH}{Casini,H. and Huerta,M. {\it Entanglement entropy for the $n$-sphere},
     \plb{694}{2010}{167}.}
   \ref{Apps}{Apps,J.S. {\it The effective action on a curved space and its conformal
     properties} PhD thesis (University of Manchester, 1996).}
   \ref{Dowcen}{Dowker,J.S., {\it Central differences, Euler numbers and symbolic methods},
 \break ArXiv:1305.0500.}
 \ref{KPSS}{Klebanov,I.R., Pufu,S.S., Sachdev,S. and Safdi,B.R.
    {\it JHEP} 1204 (2012) 074.}
 \ref{moller}{M{\o}ller,N.M. \ma {343}{2009}{35}.}
 \ref{BandO}{Branson,T., and  Oersted,B \jgp {56}{2006}{2261}.}
  \ref{BaandS}{B\"ar,C. and Schopka,S. {\it The Dirac determinant of spherical
     space forms},\break {\it Geom.Anal. and Nonlinear PDEs} (Springer, Berlin, 2003).}
 \ref{EMOT2}{Erdelyi, A., Magnus, W., Oberhettinger, F. and Tricomi, F.G. {
  \it Higher Transcendental Functions} Vol.2 (McGraw-Hill, N.Y. 1953).}
 \ref{Graham}{Graham,C.R. SIGMA {\bf 3} (2007) 121.}
  \ref{Morpurgo}{Morpurgo,C. \dmj{114}{2002}{477}.}
      \ref{DandP2}{Dowker,J.S. and Pettengill,D.F. \jpa{7}{1974}{1527}}
 \ref{Diaz}{Diaz,D.E. {\it Polyakov formulas for GJMS operators from AdS/CFT},
 {\it JHEP} {\bf 0807} (2008) 103.}
    \ref{DandD}{Diaz,D.E. and Dorn,H. {\it Partition functions and double trace
    deformations in AdS/CFT}, {\it JHEP} {\bf 0705} (2007) 46.}
    \ref{AaandD}{Aros,R. and Diaz,D.E. {\it Determinant and Weyl anomaly of
     Dirac operator: a holographic derivation}, ArXiv:1111.1463.}
  \ref{CandA}{Cappelli,A. and D'Appollonio, \pl{487B}{2000}{87}.}
  \ref{CandT2}{Copeland,E. and Toms,D.J. \cqg {3}{1986}{431}.}
   \ref{Allais}{Allais, A. {\it JHEP} {\bf 1011} (2010) 040.}
     \ref{Tseytlin}{Tseytlin,A.A. {\it On Partition function and Weyl anomaly of
     conformal higher spin fields} ArXiv:1309.0785.}
     \ref{KPS2}{Klebanov,I.R., Pufu,S.S. and Safdi,B.R. {\it JHEP} {\bf 1110} (2011) 038.}
    \ref{CaandWe}{Candelas,P. and Weinberg,S. \np{237}{1984}{397}.}
     \ref{ChandD}{Chang,P. and Dowker,J.S. \np{395}{1993}{407}.}
 \ref{Steffensen}{Steffensen,J.F. {\it Interpolation}, (Williams and Wilkins,
    Baltimore, 1927).}
     \ref{Barnesa}{Barnes,E.W. {\it Trans. Camb. Phil. Soc.} {\bf 19} (1903) 374.}
    \ref{DowGJMS}{Dowker,J.S. {\it Determinants and conformal anomalies of
    GJMS operators on spheres}, \jpa{44}{2011}{115402}.}
    \ref{Dowren}{Dowker,J.S. {\it R\'enyi entropy on spheres}, \jpamt {46}{2013}{2254}.}
 \ref{MandD}{Mansour,T. and Dowker,J.S. {\it Evaluation of spherical GJMS determinants},
 2014, Submitted for publication.}
 \ref{GandK}{Gubser,S.S and Klebanov,I.R. \np{656}{2003}{23}.}
     \ref{Dow30}{Dowker,J.S. \prD{28}{1983}{3013}.}
     \ref{Dowcmp}{Dowker,J.S. {\it Effective action on spherical domains},
      \cmp{162}{1994}{633}.}
     \ref{DowGJMSE}{Dowker,J.S. {\it Numerical evaluation of spherical GJMS operators
     for even dimensions} ArXiv:1310.0759.}
       \ref{Tseytlin2}{Tseytlin,A.A. \np{877}{2013}{632}.}
   \ref{Tseytlin}{Tseytlin,A.A. \np{877}{2013}{598}.}
  \ref{Dowma}{Dowker,J.S. {\it Calculation of the multiplicative anomaly} ArXiv: 1412.0549.}
  \ref{CandH}{Camporesi,R. and Higuchi,A. {\it J.Geom. and Physics}
  {\bf 15} (1994) 57.}
  \ref{Allen}{Allen,B. \np{226}{1983}{228}.}
  \ref{Dowdgjms}{Dowker,J.S. \jpamt{48}{2015}{125401}.}
  \ref{Dowsphgjms}{Dowker,J.S. {\it Numerical evaluation of spherical GJMS determinants
  for even dimensions}, ArXiv:1310.0759.}

\end{putreferences}

\bye